\documentclass[pra,showpacs,twocolumn]{revtex4-1}
\usepackage{slashed}
\usepackage{mathrsfs}
\usepackage{amsfonts}
\usepackage{amsmath}
\usepackage{amssymb}
\usepackage{revsymb}
\usepackage{graphicx}
\usepackage{mathrsfs}
\usepackage{bm}
\usepackage{psfrag}
\usepackage{color}
\usepackage{natbib}

\usepackage[usenames,dvipsnames]{xcolor}

\newcommand{\figrefa}[1]{Fig. \ref{#1}a}
\newcommand{\figrefb}[1]{Fig. \ref{#1}b}
\newcommand{\bea}{\begin{eqnarray}} 
\newcommand{\eea}{\end{eqnarray}} 
\newcommand{\mbf}[1]{\mathbf{#1}}
\newcommand{\trm}[1]{\textrm{#1}}

\newcommand{\Ai}{\trm{Ai}}
\newcommand{\eps}{\varepsilon}
\newcommand{\vkap}{\varkappa}
\newcommand{\vphi}{\varphi}
\newcommand{\defto}{:=}
\newcommand{\oR}{\overline{R}}
\newcommand{\oI}{\overline{\mathcal{I}}}
\newcommand{\sk}{\slashed{\varkappa}}
\newcommand{\sA}{\slashed{A}}
\newcommand{\psibar}{\overline{\psi}}

\newcommand{\av}[1]{\langle #1 \rangle}
\newcommand{\figref}[1]{Fig. \ref{#1}}
\newcommand{\figrefs}[2]{Figs. \ref{#1}, \ref{#2}}
\newcommand{\sxnref}[1]{Sec. \ref{#1}}
\newcommand{\sxnreft}[2]{Secs. \ref{#1}--\ref{#2}}
\newcommand{\eqnref}[1]{Eq. (\ref{#1})}

\newcommand{\tr}{\trm{tr}\,}
\newcommand{\Ecr}{E_{\trm{cr}}}

\fussy

\fussy

\begin{document}
\title{Photon polarisation in electron-seeded pair-creation cascades}
\author{B. \surname{King}}
  \email{ben.king@physik.uni-muenchen.de}
\affiliation{Ludwig-Maximilians-Universit\"at M\"unchen,
    Theresienstra\ss e 37, 80333 M\"unchen, Germany}
\author{N. \surname{Elkina}}
  \email{nina.elkina@physik.uni-muenchen.de}
\affiliation{Ludwig-Maximilians-Universit\"at M\"unchen,
    Theresienstra\ss e 37, 80333 M\"unchen, Germany}
\author{H. \surname{Ruhl}}
  \email{hartmut.ruhl@physik.uni-muenchen.de}
  \affiliation{Ludwig-Maximilians-Universit\"at M\"unchen,
    Theresienstra\ss e 37, 80333 M\"unchen, Germany}

\date{\today}
\begin{abstract}
An electromagnetic pair-creation cascade seeded by an electron or a photon in an
intense plane wave interacts in a complicated way with the external field. Many
simulations neglect the vector nature of photons by including their interaction
using unpolarised cross-sections. After deriving rates for the tree-level
processes of nonlinear Compton scattering and pair creation with an arbitrary linearly-polarised
photon in a constant-crossed field, we present results of
numerical simulations that include the photon's vector nature.
The simulations of seed electrons in a rotating
electric field of optical frequency on the one hand support the approximation
of using unpolarised cross-sections for tree-level processes, which predicts the same 
number of created particles when using polarised cross-section to within around
$5\%$. On the other hand, these simulations show that when the polarisation of 
the photon can be influenced by its environment, the asymmetry in the 
polarisation distribution could be used to significantly increase the rates of
each process.
\end{abstract}

\maketitle

\section{Introduction}
There are many examples of macroscopic phenomena originating from a repeated series of microscopic events. One prominent example is the process of nuclear fission, where a seed neutron collides with a $^{235}$U nucleus, releasing $^{92}$Kr, $^{141}$Ba daughter nuclei, gamma-photons and other high-energy neutrons that can further propagate a chain reaction \cite{bohr39}. Another is so-called ``particle showers'', often used by calorimeters for detection in particle physics, where an incident high-energy particle is brought to radiate, e.g. by passing through matter,  and the radiation liberates other particles which in turn can radiate and further propagate the shower \cite{blackett33,heitler60}. This can occur irrespective of whether the seed particle is charged, such as in the common case of electrons, or whether it is neutral, as in the case of photons. In such examples, it is typically a safe assumption that in the time between radiating or freeing other particles, the seed particles propagate in a simple way 
\cite{bethe34}. In contrast, the cascades of pair-creation and Compton-scattering events initiated in intense electromagnetic fields that can lead to the generation of electron-positron plasmas have a much more complicated development. From the moment the initial particles are created, using seeds or directly from vacuum, their exponential growth and recycling of the external field through absorption and re-emission can lead to a complicated interplay between the driving external field and the driven plasma, with such systems predicted to occur, for example when intense electromagnetic fields irradiate single or collections of particles \cite{ruhl10, elkina11}, for example in solids \cite{ridgers12}. Although electron-seeded pair creation has been demonstrated experimentally \cite{burke97}, profuse positron creation with lasers has thus far been mainly demonstrated via the Bethe-Heitler process of pair-creation by a high-energy photon in the Coulomb field of a nucleus \cite{bethe34, chen09}. 
\newline

In order to better understand such systems, there has been an intensification of research efforts to simulate such plasmas \cite{ruhl10, elkina11, ridgers12, hebenstreit13, ilderton13c}. Due to their complexity, to model a large number of particles, many approximations have to be made. The purpose of this paper is to investigate one such approximation, namely that the polarisation of photons propagating the cascade can be effectively neglected, being set to the average polarisation angle for each tree-level process. To achieve this, we also present a derivation of the linearly-polarised Compton-scattering cross-section in a constant crossed field, which we were unable to find anywhere else in the literature, although the unpolarised cross-section and pair-creation cross-section for definite polarisation have been derived some time ago \cite{kibble64, nikishov64}. Most recently, arbitrary photon polarisation has been studied in relation to tree-level Compton scattering \cite{krajewska12a} and pair-creation \cite{krajewska12b} in finite laser pulses (a review of 
strong-field QED effects can be found in \cite{ritus85, marklund_review06, ehlotzky09, dipiazza12}). One instance where photon polarisation in relativistic plasmas is expected to play a role is in the strong magnetic field of certain astrophysical objects such as magnetars \cite{thompson95, harding06}.
\newline

In the current paper, we present calculations performed for a constant-crossed-field background as this is a good approximation when the formation lengths of processes are much smaller than inhomogeneities in the field. More precisely, any arbitrary, time-dependent background can be considered constant on the QED spacetime scale when \linebreak ${\xi = (e^{2} p_{\mu}T^{\mu\nu}p_{\nu})/(m^{2} (\vkap p)^{2}) \ll 1}$, (using the definition of $\xi$ derived in \cite{ilderton09}) where $T^{\mu\nu}$ is the energy-momentum tensor, $\vkap$ is the external-field wavevector, $p$ is the momentum of particle involved, $e>0$ and $m$ are the charge and rest-energy of a positron respectively and we work in a system of units in which $\hbar=c=1$.  In terms of laser fields, $\xi$ is often referred to as the ``intensity'' or ''classical nonlinearity''  parameter, $\xi= m\chi_{E}/\omega$, $\chi_{E} = E_{0}/\Ecr$, $E_{0}$ is the electric field amplitude and $\Ecr=m^{2}/e$ is the critical, so-called ``Schwinger'' field. Moreover,
 an arbitrary, constant field can then be expressed in terms of three relativistic invariants:
\bea
\chi = \frac{e \sqrt{|p_{\mu}F^{\mu\nu}|^{2}}}{m^{3}};~\mathcal{F}=\frac{e^{2} F_{\mu\nu}F^{\mu\nu}}{4 m^{4}};~\mathcal{G}=\frac{e^{2} F^{\ast}_{\mu\nu}F^{\mu\nu}}{4 m^{4}}, \label{eqn:paramdef}
\eea
where $F$ and $F^{\ast}$ are the electromagnetic tensor and its dual. Any function of these three parameters $W(\chi,\mathcal{F}, \mathcal{G})$ can be considered $\approx W(\chi,0,0)$, when $\mathcal{F}, \mathcal{G} \ll \chi^{2}, 1$. Such functions then describe processes in ``crossed'' fields ($\mbf{E}\cdot\mbf{B} = E^{2}-B^{2} = 0$, for electric and magnetic field $\mbf{E}$, $\mbf{B}$, equivalently $F^{2} = F^{\ast}F=0$). At least for laser systems, since $E/\Ecr \ll 1$, the second of these inequalities is easily fulfilled and as the processes in question only become probable when $\chi \gtrsim 1$, the first inequality will also be fulfilled in the current study. A pedagogical description of constant crossed field Compton scattering has recently been given in \cite{dinu12}.
\newline

The paper is organised as follows. In \sxnref{sxn:Comp} we present the derivation of Compton-scattering of a polarised photon in a constant crossed field, discuss the result, then present in \sxnref{sxn:Pair} the rate for creation of pairs due to an arbitrarily linearly-polarised photon, which is followed in \sxnref{sxn:CascTh} by a study of these two processes combined -- the smallest chain of events considered involving a real photon that can lead to $e^{-}$-seeded pair creation (the two-step trident process), and finally the conclusions of the theoretical sections are investigated in \sxnref{sxn:CascSim} where results are presented from simulations of chains of lowest-order processes to compare the effect of including polarisation in pair-creation cascades.
\newline 


\section{Polarised Compton scattering in a constant crossed field} \label{sxn:Comp}
\begin{figure}[!h]
\noindent\centering
 \includegraphics[draft=false, width=7cm]{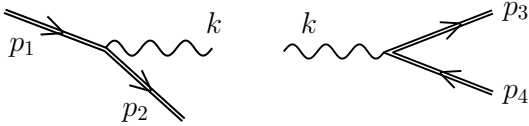}
\caption{One example generation in the envisaged cascade, with an on-shell photon linking the processes of Compton-scattering (left) and pair-creation (right). Double lines represent dressed wavefunctions that include the interaction with the external field to all orders. \label{fig:feydiag}} 
\end{figure}
The vector potential of a plane-wave external field $A^{\mu} = A^{\mu}(\varphi)$, is solely a function of the phase $\varphi = \vkap x$. The solutions to Dirac's equation in such a background field for a particle of momentum $p$ are described by the so-called Volkov wavefunctions \cite{volkov35}:
\begin{gather}
\psi_{r}(p) = \Big[1+\frac{e \sk\sA}{2\varkappa p}\Big] \frac{u_{r}(p)}{\sqrt{2p^{0}V}} \mbox{e}^{iS}; \\
 S = -px - \int^{\varphi}_{\varphi_{0}} d\varphi' \,\Big(\frac{e(pA(\varphi'))}{\vkap p} - \frac{e^{2}A^{2}(\varphi')}{2(\vkap p)}\Big),
\end{gather}
where $\psi_{r}$ are incoming fermion wavefunctions, $u_{r}(p)$ are free-electron spinors, $\slashed{A} = \gamma^{\mu}A_{\mu}$, $V$ is the system volume and $S$ corresponds to the classical action of an electron in a plane wave \cite{landau4}. The limit of a constant crossed field is achieved by choosing $A^{\mu}(\varphi) = a^{\mu}\varphi$ and letting $\varkappa^{0}\to0$ when all dependency on $\varkappa^{0}$ has disappeared. The amplitude for Compton scattering (the left-hand diagram in \figref{fig:feydiag}) is given by:
\bea
S_{fi,\gamma} &=& e \int\!d^{4}x \,\, \psibar_{2}(x)\slashed{\eps} \frac{\mbox{e}^{ikx}}{\sqrt{2k^{0}V}}\psi_{1}(x),
\eea
where $k$ is the momentum of the real photon and $\psi_{j}=\psi(p_{j})$ with spinor indices suppressed. Employing the constant-crossed-field limit, one can write this as:
\begin{gather}
S_{fi,\gamma} = e \int\!d^{4}x \,\,\mbox{e}^{i(p_{2}+k-p_{1})x} F(\varphi) \\ 
F(\varphi) = \mbox{e}^{i\Phi(b_{2},b_{3})} \frac{\overline{u}_{r}(p_{2})}{\sqrt{2p_{2}^{0}V}}\Big[1+\frac{e \sA\sk}{2\varkappa p_{2}}\Big] \slashed{\eps} \Big[1+\frac{e \sk\sA}{2\varkappa p_{1}}\Big] \frac{u_{r}(p_{1})}{\sqrt{2p_{1}^{0}V}}\nonumber\\
b_{2} = -\frac{e}{2} \left(\frac{p_{1}a}{\vkap p_{1}} - \frac{p_{2}a}{\vkap p_{2}} \right);\qquad b_{3} = \frac{e^{2}a^{2}}{6} \left(\frac{1}{\vkap p_{1}} - \frac{1}{\vkap p_{2}} \right),
\end{gather}
where $\Phi(b_{2},b_{3})=b_{2}\varphi^{2}+b_{3}\varphi^{3}$. By Fourier-transforming $F(\varphi)$ and integrating over $x$ one acquires
\begin{gather}
S_{fi,\gamma} = (2\pi)^{3}e \int\!dr~\delta^{4}(p_{2}+k-p_{1}-r\vkap)~\Gamma(r) \label{eqn:Sfi1} \\ 
\Gamma(r) = \int\!d\varphi~ F(\varphi)\,\mbox{e}^{ir\varphi}. 
\end{gather}
To obtain the polarised rate of Compton scattering, $R_{\gamma}$, we use 
\bea 
R_{\gamma} = \frac{V^{2}}{2T}\int\!\!\frac{d^{3}p_{2}}{(2\pi)^{3}}\,\frac{d^{3}k}{(2\pi)^{3}}~\tr|S_{fi,\gamma}|^{2}, \label{eqn:Rgam}
\eea
where $\tr$ is the trace over spin indices, the factor $1/2$ is due to an average over initial electron spin states and $T$ is the system duration. Using lightfront co-ordinates for momenta $p^{+,-} = (p^{0}\pm p^{3})/2$, $p^{\perp} = (p^{1}, p^{2})$ and for co-ordinates $x^{+,-} = x^{0}\pm x^{3}$, $x^{\perp} = (x^{1}, x^{2})$ and defining at this point a specific co-ordinate system for calculations $\vkap = \vkap^{0}(1, 0, 0, 1)$, $a_{1} = (0,1,0,0)$ and $a_{2}=(0,0,1,0)$, we use the following arguments to deal with the delta-function in \eqnref{eqn:Sfi1}:
\begin{widetext}
\bea
|S_{fi,\gamma}|^{2} &=& (2\pi)^{6}e^{2}\int\!dr\,dr'~\delta^{4}(\Delta p-r\vkap)\delta^{4}(\Delta p-r'\vkap)\Gamma(r)\Gamma^{\dagger}(r')\\
|S_{fi,\gamma}|^{2} &=& (2\pi)^{6}e^{2}\int\!dr\,dr'~\delta^{4}(\Delta p-r\vkap)\frac{\delta^{4}[(r'-r)\vkap]}{\delta(r-r')}\delta(r-r')\Gamma(r)\Gamma^{\dagger}(r')\\
|S_{fi,\gamma}|^{2} &=& (2\pi)^{6}e^{2}\frac{VT}{(2\pi)^{3}L_{\varphi,\gamma}}\int\!dr~\delta^{4}(\Delta p-r\vkap)|\Gamma(r)|^{2}\\
|S_{fi,\gamma}|^{2} &=& (2\pi)^{3}e^{2}\frac{VT}{L_{\varphi,\gamma}\vkap^{0}}\delta^{(2)}(\Delta p^{\perp})\delta(\Delta p^{-})|\Gamma(r_{\ast})|^{2},
\eea
\end{widetext}
where we have defined $\Delta p = p_{2}+k -p_{1}$ and in the final line integrated over the $^{+}$ component of the delta-function to give:
\bea
\vkap^{0}r_{\ast} = \Delta p^{+} = \frac{\vkap^{0}}{2p_{1}\vkap} ((p_{2}+k)^{2}-m^{2})= \vkap^{0}\frac{p_{2}k}{p_{1}\vkap},
\eea
and used the on-shell property of momenta, where we have defined a dimensionless interaction phase length $L_{\varphi,\gamma}$, following standard arguments in e.g. \cite{ritus85}:
\bea
\delta(r-r')\big|_{r=r'} = \int \!\frac{dl}{2\pi}~ \mbox{e}^{i(r-r')l}\Big|_{r=r'} =  \frac{L_{\varphi,\gamma}}{2\pi}. \label{eqn:Ldef}
\eea
By noting that:
\bea
F_{n}(r,b_{2},b_{3}) &\defto& \int_{-\infty}^{\infty} d\vphi~ (i\vphi)^{n-1} \mbox{e}^{i ( r\vphi + b_{2} \vphi^{2} + b_{3}\vphi^{3})}, \label{eqn:Airydef1}
\eea
can be written in terms of the Airy function $\Ai$ \cite{olver97} and its derivative, $\Ai'$, where, for example:
\begin{gather}
F_{1} = f_{1}\,\Ai(\nu)\label{eqn:I0a};\qquad f_{1} =  \frac{2\pi\mbox{e}^{i\eta}}{(3b_{3})^{1/3}};\nonumber\\
\eta = -\frac{rb_{2}}{3b_{3}} +\frac{2b_{2}^{3}}{27b_{3}^{2}}; \qquad \nu= \frac{r-b_{2}^{2}/3b_{3}}{(3b_{3})^{1/3}}, \label{eqn:Fdefs}
\end{gather}
and $F_{n} = \partial^{n-1}\,F_{1}/\partial r^{n-1}$ for $n\in \mathbb{N}_{>0}$, performing the spin trace of $\Gamma(r)$, one arrives at:
\bea
\frac{\tr|S_{fi,\gamma}|^{2}}{VT} &=& \frac{\pi^{3}e^{2}\delta^{(2)}(\Delta
p^{\perp})\delta(\Delta p^{-})}{p^{0}_{1}\,p^{0}_{2}\,
k^{0}L_{\varphi,\gamma}\vkap^{0}}\tr
|\Gamma(r_{\ast})|^{2},\nonumber\\
\frac{1}{8}\tr|\Gamma(r_{\ast})|^{2} &=& \Big|p_{1}\eps'^{\ast}\,F_{1}
-ie\,a\eps'^{\ast}\,F_{2}\Big|^{2} \nonumber \\
&&+\frac{3b_{3}}{2}\, \vkap
k\,\Big(|F_{2}|^{2}+\trm{Re}\,F_{1}F_{3}\Big)
\label{eqn:ritAgree}
\eea
where $\eps'$ is related to the photon polarisation $\eps$ via $\eps'^{\mu}\defto \eps^{\mu}-k^{\mu} (\vkap \eps)/(\vkap k)$ and \eqnref{eqn:ritAgree} agrees with \cite{ritus85} (P. 557, Eq. (36)). The redefinition of $\eps'$ inspired by
\cite{ritus85} is also a valid polarisation vector, obeying $\eps'^{2}=-1$
and $\eps' k = 0$ as required, but is useful in removing higher powers of
$k_{x,y}$ from the spin trace. Let us use the following basis for the two
polarisation vectors transverse to the photon wavevector (e.g. as used in
\cite{baier75a}):
\bea
\Lambda_{1,2}^{\mu} = \frac{k\vkap ~a_{1,2}^{\mu} - ka_{1,2}\, \vkap^{\mu}}{\vkap k ~\sqrt{-a_{1,2}^{2}}}, \quad a_{i}a_{j} = -\delta_{ij}\left(\frac{E}{\vkap_{0}}\right)^{2}\!\!, \label{eqn:Lambdas}
\eea
where $E$ is the modulus of the electric field, then $\Lambda^{\mu}_{i}\Lambda_{j,\mu} = -\delta_{i,j}$ and $\Lambda_{i} k = 0$ for $i, j, \in \{1,2\}$ as required. When $\vkap k=0$ the rate vanishes quicker than $1/\vkap k$, so the definition in \eqnref{eqn:Lambdas} is sound (see also \cite{dinu12} for an analysis of collinear divergences in Compton scattering). So for a head-on collision of photon and external-field wave-vector, $\Lambda_{1,2} = a_{1,2}/(-a_{1,2}^{2})^{1/2}$. This basis can also be written in terms of the difference of incident and outgoing fermion momenta, by defining $\delta p = p_{1}-p_{2}$ to give:
\bea
\Lambda_{1,2}^{\mu} = \frac{\vkap\delta p ~a_{1,2}^{\mu} - a_{1,2}\delta p\, \vkap^{\mu}}{\vkap \delta p ~\sqrt{-a_{1,2}^{2}}},
\eea
where on average, the angle between $p_{2}$ and $p_{1}$ becomes smaller the more relativistic $p_{1}$ is.
We seek the rate of scattering for arbitrary linear polarisation. To this end, define the polarisation to be a superposition of these basis vectors
\bea
\eps^{\mu} = c_{1} \Lambda_{1}^{\mu} + c_{2} \Lambda_{2}^{\mu},\qquad c_{1},c_{2} \in \mathbb{C}. 
\eea
Since $\eps^{2} = -1$, we know $c_{2}^{2} = 1-c_{1}^{2}$. When one combines the expression for the rate $R_{\gamma}$ in \eqnref{eqn:Rgam} with \eqnref{eqn:ritAgree} and integrates the delta-functions over $p_{2}$, just as for the unpolarised cross-section, the integrand is independent of $k_{x}$. However, making the observation $\int\,dk_{x} = m^{2}\chi_{E}\chi_{k}/(\vkap^{0}\chi_{1}) \int d\varphi_{\ast}$, where $\varphi_{\ast}$ is the saddle-point of the Airy functions in the problem \eqnref{eqn:Airydef1}, and noting that this is the same interaction phase length $L_{\varphi,\gamma}$ defined in \eqnref{eqn:Ldef}, the integral can be performed, cancelling the $L_{\varphi,\gamma}$ factors. The final manageable integral in $k_{y}$ is then calculated using Airy integral identities given in \cite{king13b, aspnes66}. One then arrives at the rate for Compton scattering for an arbitrarily linearly-polarised photon emitted in a constant-crossed field:
\begin{widetext}
\begin{gather}
R_{\gamma}(\phi) = \frac{-\alpha m^{2}}{p_{1}^{0}}\int_{v_{\trm{min}}}^{\infty} \frac{dv}{(1+v)^{2}} \left\{\frac{1}{z} \left[2\cos^{2}\phi+1 + \frac{v^{2}}{1+v} \right]\Ai'(z) + \Ai_{1}(z)\right\}, \label{eqn:Rg}\\
\phi \in [0,\pi[;\qquad z=\mu^{2/3};\qquad  \mu = \frac{\chi_{k}}{\chi_{1}(\chi_{1}-\chi_{k})}=\frac{v}{\chi_{1}};\qquad \chi_{k} = \frac{e\sqrt{|F_{\mu\nu}k^{\nu}|^{2}}}{m^{3}} = \frac{2\chi_{E}k^{-}}{m},
\end{gather}
\end{widetext}
where we have defined a polarisation angle $c_{1} = \cos\phi$, $c_{2} = \sin\phi$, $\alpha=e^{2}/4\pi$ is the fine-structure constant and $v_{\trm{min}}\geq0$ permits a photon momentum cutoff. The polarisation angle $\phi$ is then the angle of photon polarisation in the photon's transverse plane (the angle to basis vector $\Lambda_{1}$). As a relativistic electron radiates in a cone of angle $\sim1/\gamma$ around its momentum vector \cite{jackson99} ($\gamma =
(1-(v/c)^{2})^{-1/2}$ where $v$ is the particle velocity) (see also e.g. \cite{harvey09}), for a head-on collision of electron and external field wavevector, $\phi=0,\pi/2$ correspond approximately to the $1$- and $2$-directions. \\

As a test of the polarised rate $R_{\gamma}(\phi)$ in \eqnref{eqn:Rg} we note that if one defines $\overline{R}_{\gamma} = [R_{\gamma}(0)+R_{\gamma}(\pi/2)]/2=R_{\gamma}(\pi/4)$ as the Compton scattering rate averaged over polarisation states, the so-called ``unpolarised'' rate, then $\overline{R}_{\gamma}$ can be seen to 
agree with other results in the literature e.g. \cite{ritus85} (P. 559, Eq. (49)). One can explain the polarisation dependence of $R_{\gamma}(\phi)$. Since the polarisation vector is normalised, and physical observables depend upon the scalar product of this vector with others occurring in the problem, the projections $\cos\phi$, $\sin\phi$ onto the basis vectors could have been seen to appear. As probabilities depend upon the square of these variables and as the external field is homogeneous the resulting rates could have been seen to depend on $2\phi$, restricting the range of $\phi$ to $\phi\in[0,\pi[$. The basis chosen is useful as $\Lambda_{1}a = -1$, $\Lambda_{2}a=0$ and $\Lambda_{1,2}\vkap=0$. In a constant crossed field, we expect the final rate to depend upon the quantum non-linearity parameter $\chi$. Since the definition of $\chi$ is symmetric in electric-field vector i.e. $a\to-a$, $c_{1}\to-c_{1}$ or equivalently $\phi \to \pi-\phi$ should be a further symmetry, allowing one to further curtail the important range of $\phi$ to $\phi \in [0,\pi/2[$. It follows that the dependency on polarisation angle in the final rate must be of the form $\cos^{2}\phi$, leading to the result that $\av{R_{\gamma}(\phi)}_{\phi}=R_{\gamma}(\av{\phi}_{\phi})$. If other basis vectors were chosen or if the field were not homogeneous or constant, this would not necessarily be the case.\\

\begin{figure}[!h]
\centering\noindent
 \includegraphics[draft=false,width=7cm]{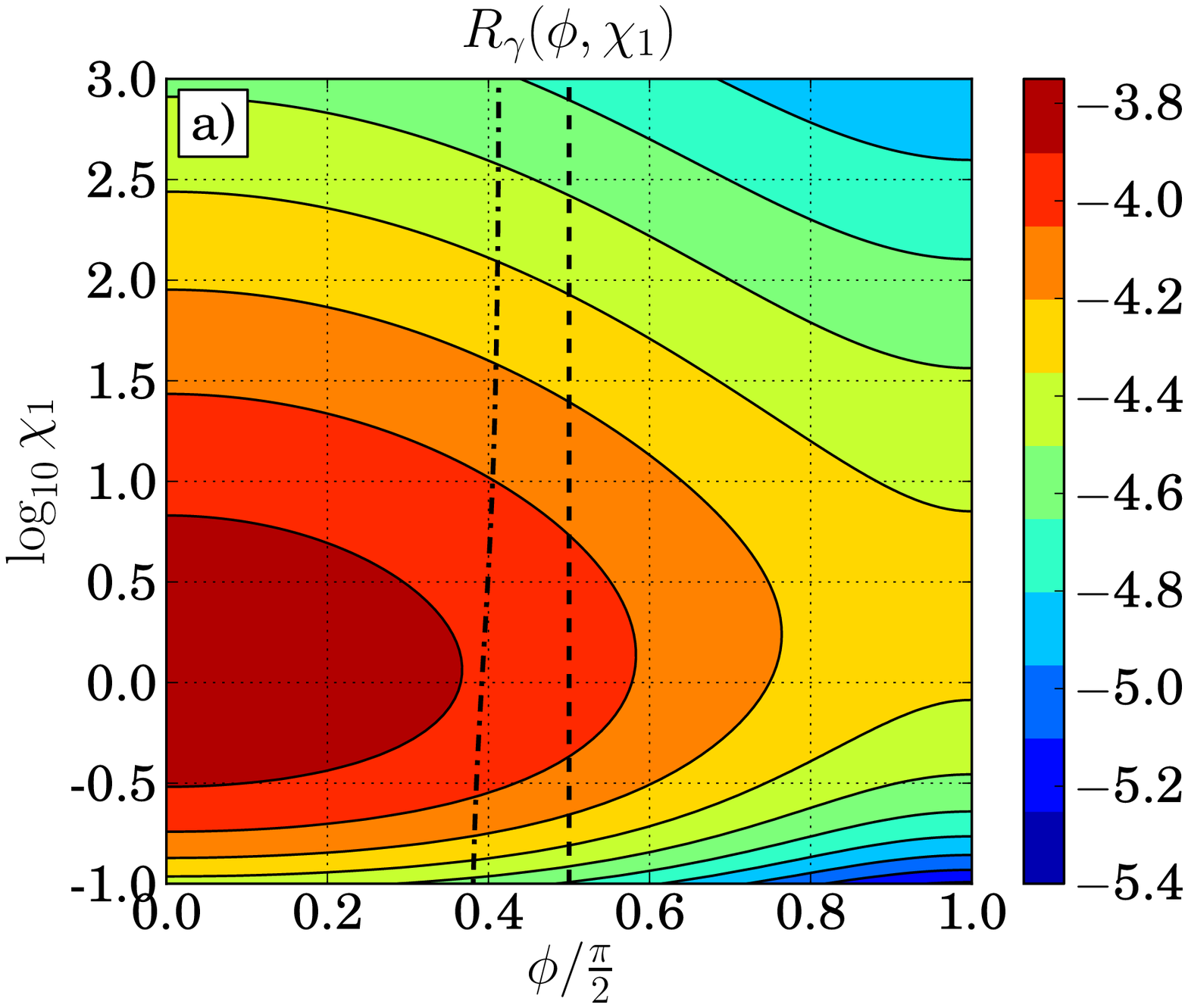} \\
\includegraphics[draft=false,width=7cm]{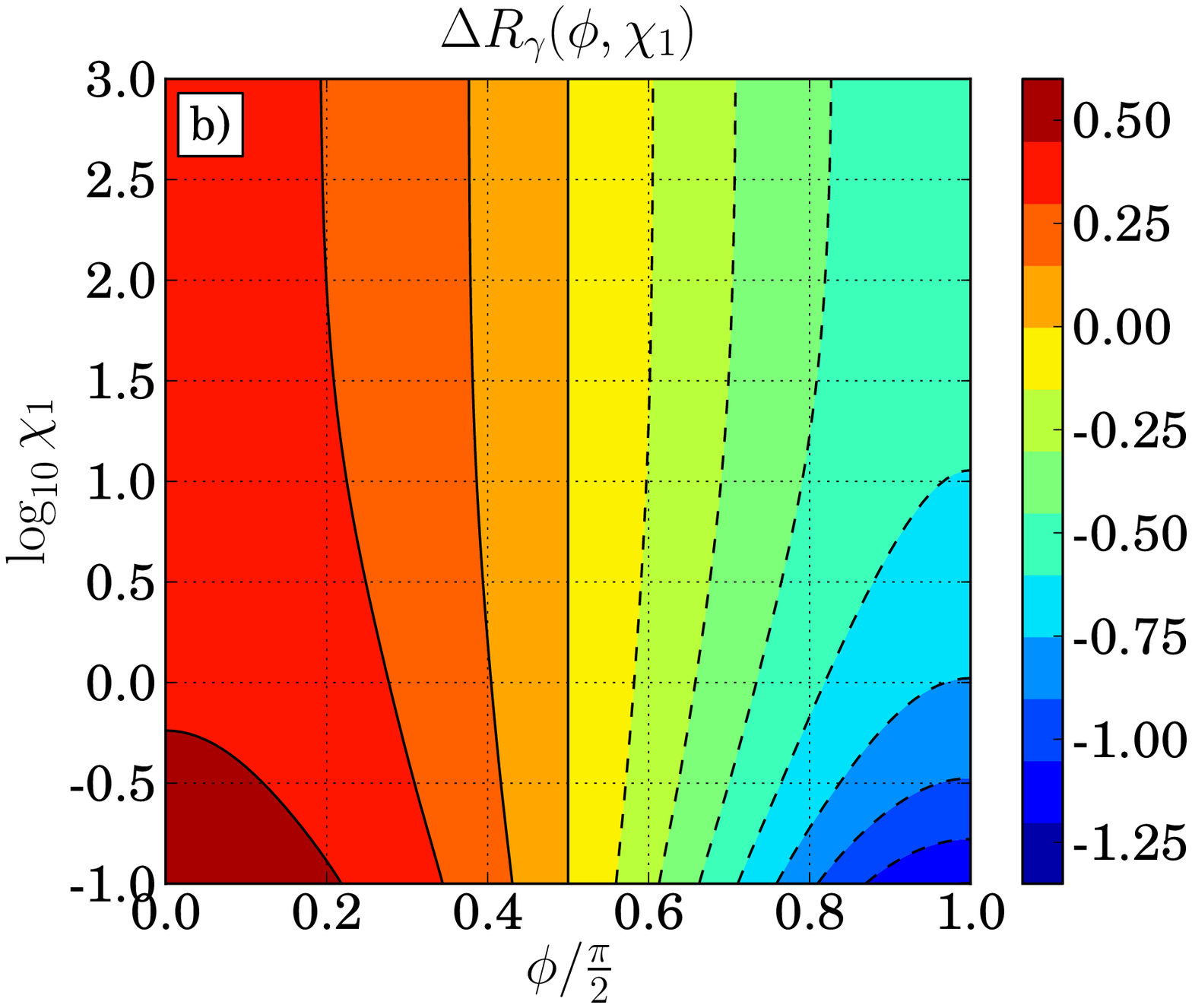}
\caption{(Color online). 
The rate for Compton scattering with momentum cutoff $\chi_{k}\geq 0.01$ and head-on collision of electron and external field wave-vector with $\chi_{E}=0.01$. In plot a) the dashed line indicates the unpolarised Compton scattering rate $\oR_{\gamma}$, equivalent to taking the average over polarisations whereas the dot-dashed line indicates the average polarisation produced. In plot b) is the relative difference from taking the unpolarised rate $\Delta R_{\gamma} = 2(R_{\gamma}(\phi)-\oR_{\gamma})/(R_{\gamma}(\phi)+\oR_{\gamma})$. 
}\label{fig:Rgs}
\end{figure}

Comparison of the asymptotic limits for the unpolarised non-linear Compton scattering rate given in \cite{ritus85} and the full polarised rate $R_{\gamma}(\phi)$ yielded the following lowest-order asymptotic limits:
\begin{equation}
R_{\gamma}(\phi) \sim 
\begin{cases}
\frac{\alpha}{\sqrt{3}}\frac{\chi_{1}}{p_{1}^{0}}(1-\frac{8\sqrt{3}}{15}\chi_{1})(3\cos^{2}\phi+1) & \chi_{1} \ll 1\\
\frac{4\alpha\Gamma(2/3)}{27}\frac{(3\chi_{1})^{2/3}}{p_{1}^{0}}(3\cos^{2}\phi+2) & \chi_{1} \gg 1.
\end{cases}
\end{equation}
By considering the $\phi$-dependent factors in the asymptotic limits, one can acquire the expected polarisation of photon produced in these limits, $\trm{E}[\phi_{\gamma}]$:
\begin{equation}
\frac{2}{\pi}\trm{E}[\phi_{\gamma}] \sim
\begin{cases}
\frac{1}{2}-\frac{6}{5\pi^{2}} \approx 0.378 & \chi_{1} \ll 1\\
\frac{1}{2}-\frac{6}{7\pi^{2}} \approx 0.413& \chi_{1} \gg 1,
\end{cases}
\end{equation}
which we note is around $10\%$ lower than the average polarisation $\av{\phi}=\pi/4$ (indicated by the dot-dashed line in \figrefa{fig:Rgs}).

In \figrefa{fig:Rgs}, $R_{\gamma}(\phi)$ is plotted as a function of incoming quantum nonlinearity parameter $\chi_{1}$ as well as the polarisation angle $\phi$, for the photon momentum cutoff $\chi_{k}\geq 0.01$. We define the relative difference from the unpolarised rate, $\Delta R_{\gamma}=2(R_{\gamma}(\phi)-\oR_{\gamma})/(R_{\gamma}(\phi)+\oR_{\gamma})$, plotted in \figrefb{fig:Rgs}. Although the relative difference is largest for small $\chi_{1}$, we note for the optimum region around $\chi\approx1$, there still persists a maximum relative difference $\Delta R_{\gamma}$ of around $+35\%$, $-65\%$.

\section{Polarised pair creation in a constant crossed field} \label{sxn:Pair}
Pair-creation (the right-hand diagram in \figref{fig:feydiag}) is a cross-channel of Compton scattering, which can be arrived at by making the substitution \cite{landau4} $p_{1}\to -p_{4}$, $p_{2}\to p_{3}$, $k\to -k$ in \eqnref{eqn:ritAgree}, for outgoing electron and positron  momenta $p_{3}$ and $p_{4}$. By following similar steps to the Compton-scattering derivation, one acquires:
\begin{widetext}
\begin{gather}
R_{e}(\phi) = \frac{-\alpha m^{2}}{2k^{0}}\int_{1}^{\infty} \frac{du}{u\sqrt{u(u-1)}} \left\{\frac{1}{z} \left[4u-1-2\cos^{2}\phi \right]\Ai'(z) - \Ai_{1}(z)\right\}, \label{eqn:Re}\\
\phi \in [0,\pi[;\qquad z=\mu^{2/3};\qquad  \mu = \frac{\chi_{k}}{(\chi_{k}-\chi_{3})}=\frac{4u}{\chi_{k}};\qquad \chi_{3} = \frac{e\sqrt{|F_{\mu\nu}p_{1}^{\nu}|^{2}}}{m^{3}} = \frac{2\chi_{E}p_{3}^{-}}{m}.
\end{gather}
\end{widetext}
This expression can be tested in an even clearer way than $R_{\gamma}(\phi)$, by comparing $R_{e}(0)$ and $R_{e}(\pi/2)$ rates with the known rates for these polarisations, which exactly reproduce the expressions given in e.g. \cite{ritus85}. We again define the unpolarised rate as $\oR_{e}=[R_{e}(0)+R_{e}(\pi/2)]/2=R_{e}(\pi/4)$. \\

The asymptotic limits for pair-creation in a constant crossed field by a polarised photon take the form:
\begin{equation}
R_{e}(\phi) \sim 
\begin{cases}
\frac{\alpha\sqrt{3}}{8}\frac{\chi_{k}}{k^{0}}\mbox{e}^{-8/3\chi_{k}}(2-\cos^{2}\phi) & \chi_{k} \ll 1\\
\frac{3\alpha[\Gamma(2/3)]^{4}}{14\pi^{2}}\frac{(3\chi_{k})^{2/3}}{k^{0}}(3-\cos^{2}\phi) & \chi_{k} \gg 1.
\end{cases}
\end{equation}
Again, one can calculate the expected polarisation of photon leading to pair creation, $\trm{E}[\phi_{e}]$, yielding:
\begin{equation}
\frac{2}{\pi}\trm{E}[\phi_{e}] \sim
\begin{cases}
\frac{1}{2}+\frac{2}{3\pi^{2}} \approx 0.567 & \chi_{1} \ll 1\\
\frac{1}{2}+\frac{2}{5\pi^{2}} \approx 0.541& \chi_{1} \gg 1.
\end{cases}
\end{equation}
\begin{figure}[!h]
\centering\noindent
 \includegraphics[draft=false, width=6.93cm]{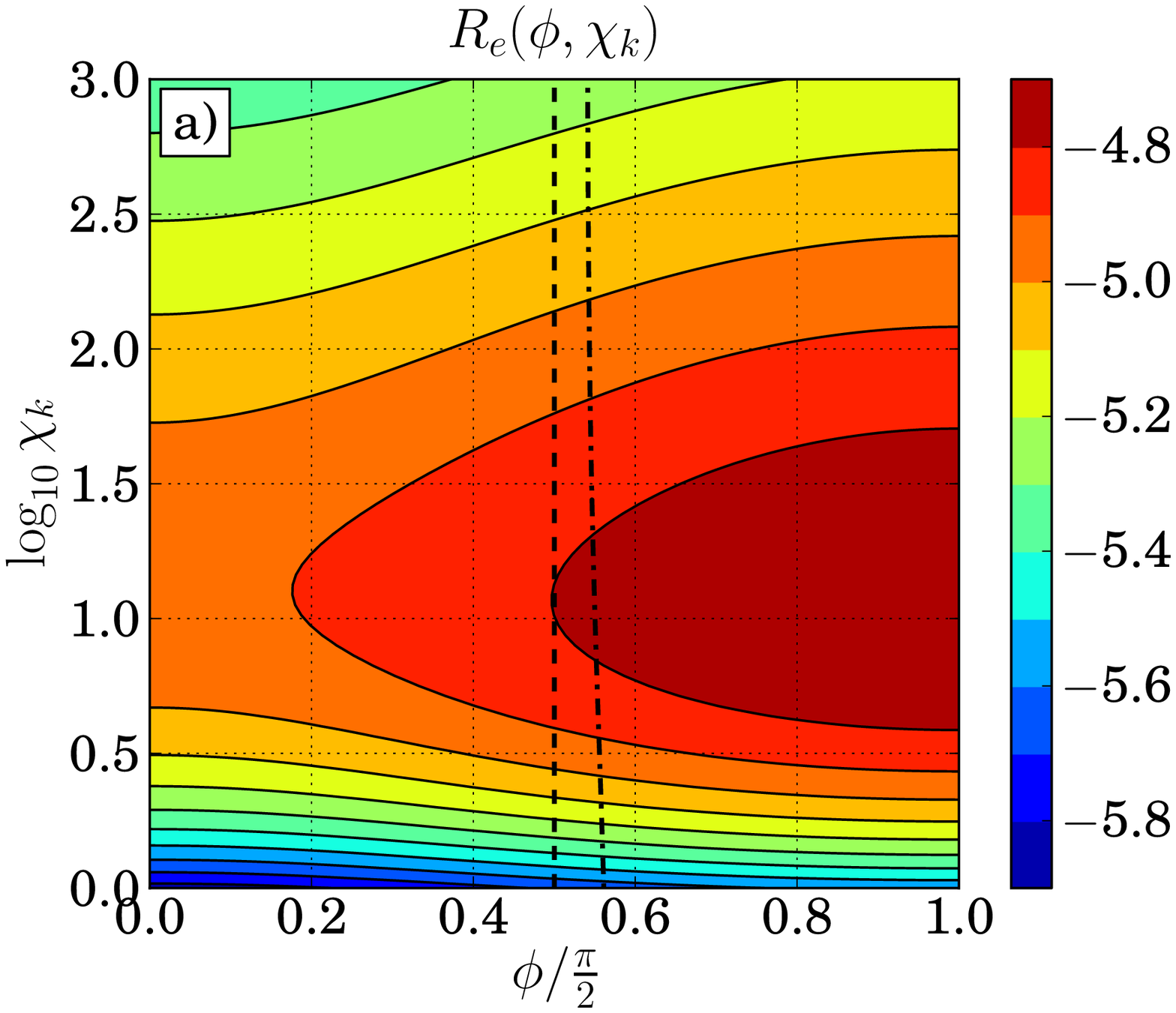} \\
 \includegraphics[draft=false, width=6.93cm]{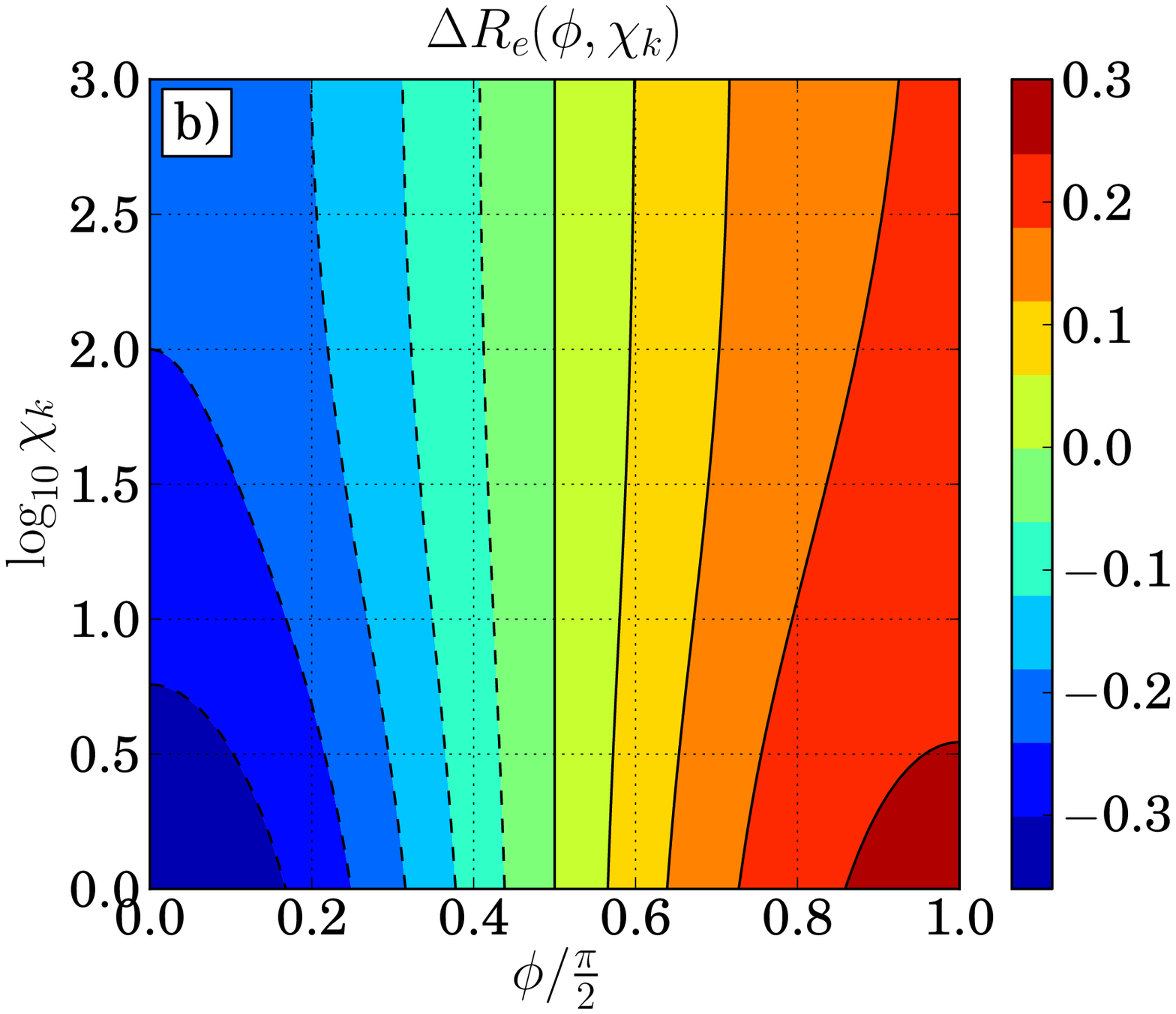}
\caption{(Color online). Pair-creation rates for a photon propagating antiparallel to the field ($\chi_{E}=0.01$). In plot a), the dashed line traces the unpolarised pair-creation rate $\oR_{\gamma}$ and the dot-dashed line the average polarisation leading to pair-creation. Plot b) is of the relative difference from taking the unpolarised rate $\Delta R_{e} = 2(R_{e}(\phi)-\oR_{e})/(R_{e}(\phi)+\oR_{e})$.} \label{fig:Res}
\end{figure}
In \figref{fig:Res} we plot how the pair-creation rate depends on $\chi_{k}$ and polarisation angle as well as the relative difference due to polarisation. We note that the optimum rate for pair-creation is at a typically higher value of the quantum non-linearity parameter than for Compton scattering, $\chi_{k} \approx 10^{1.1}$. Also, the plot of $\Delta R_{e}$ shows that photon polarisations which are \emph{more} likely to be produced via nonlinear Compton scattering are \emph{less} likely to lead to pair-creation and vice-versa. Due to the different shapes of $R_{e}$ and $R_{\gamma}$, we will further investigate in the next section whether this compensation is seen in a cascade.

\section{Photon polarisation in two-step fermion-seeded pair creation} \label{sxn:CascTh}
Electron- (positron-) seeded pair creation in an external field $e^{\pm}\to e^{\pm}+e^{+}e^{-}$ can proceed via a two-step process, where the intermediate photon becomes real and then decays into a pair ($e^{\pm}\to e^{\pm}+\gamma$, $\gamma \to e^{+}e^{-}$), or via a one-step process where the intermediate photon remains virtual \cite{ilderton11}. Until now, it has been shown that the one-step process can become dominant by tuning the external-field frequency  to exploit a resonance in the photon propagator in the multi-photon regime ($\xi \ll 1 $) \cite{hu10}. As the constant-crossed-field calculation is valid in the limit of zero external-field frequency ($\xi \to \infty$), this effect can be ruled out. It has also been suggested that the one-step process can become important for length-scales $\lambdabar_{\ast}=\lambdabar/\chi_{E}$, where $\lambdabar=1/m$ is the reduced Compton wavelength \cite{king13b}. Therefore we restrict our analysis to scales $L\gg \lambdabar_{\ast}$, which is also the condition for the constant crossed field to be a valid approximation to an 
arbitrary external field. Assuming that spin effects of the incoming and outgoing fermions are negligible (including spin-effects originating from radiative corrections to the Volkov states \cite{meuren11}), it is supposed that taking fermion-seeded pair creation in a constant crossed field to be given entirely by the two-step process, is a good approximation. Let us write this in terms of the probability $P_{\gamma e}$ of fermions to seed the two-step process, in a formation length $L_{\varphi}$. To do this, we use the relation $T_{j}/p_{j}^{0} = L_{\varphi,j}/(\vkap p_{j})$, where $j\in \{\gamma, e\}$ (where $p_{\gamma}=k$ and $p_{e}=p_{1}$). Then $P_{\gamma e}=L^{2}_{\varphi} \mathcal{I}_{\gamma e}$, where $\mathcal{I}_{\gamma e}$ is the dynamical part of the rate, given by:
\bea
\mathcal{I}_{\gamma e}(\phi) = \int \!dv~\frac{\partial R_{\gamma}(v,\phi)}{\partial v}\,R_{e}\left(\chi_{k}(v),\phi\right) \frac{k^{0}p^{0}_{1}/m^{2}}{\chi_{1}\chi_{k}(v)}, \label{eqn:twostep}
\eea
and $\chi_{k}(v)=\chi_{1}v/(1+v)$. If the locally-constant-field approximation is employed, the factor $L^{2}_{\varphi}$ can be understood as a double integration over external-field phases $\varphi$ as $L^{2}_{\varphi}\mathcal{I}_{\gamma e}(\varphi_{\gamma},\varphi_{e}) = (m\chi_{E}/\varkappa^{0})^{2} \int_{-\infty}^{\varphi}d\varphi_{\gamma}\int^{\varphi_{\gamma}}_{-\infty} d\varphi_{e}\,\mathcal{I}_{\gamma e}(\varphi_{\gamma},\varphi_{e})$. \\

\begin{figure}[!h]
\centering\noindent
\includegraphics[draft=false,width=7cm]{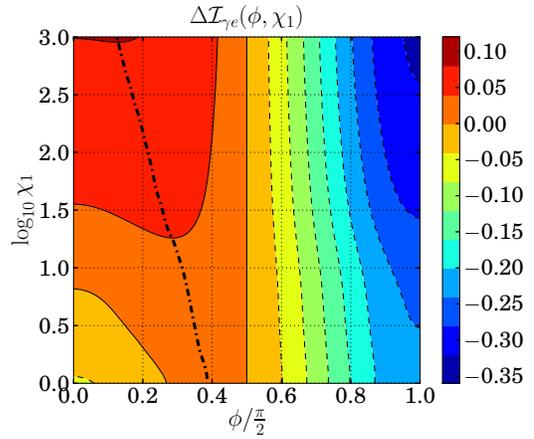}
\caption{(Color online). A plot of the relative difference in the dynamical part of the rate for the two-step fermion-seeded pair creation using a polarised intermediate photon compared to using unpolarised rates $\Delta \mathcal{I}_{\gamma e} = 2(\mathcal{I}_{\gamma e}(\phi)-\overline{\mathcal{I}}_{\gamma e})/(\mathcal{I}_{\gamma e}(\phi)+\overline{\mathcal{I}}_{\gamma e})$, for a head-on collision of electron and external field wave-vector with $\chi_{E}=0.01$. The dot-dashed line traces the polarisation of photon most likely to facilitate the two-step process.}\label{fig:DelRge}
\end{figure}

The dependency of two-step fermion-seeded pair creation on the intermediate photon polarisation is shown in \figref{fig:DelRge}, displaying the relative difference to using unpolarised rates for each sub-process $\Delta \mathcal{I}_{\gamma e} = 2(\mathcal{I}_{\gamma e}(\phi)-\overline{\mathcal{I}}_{\gamma e})/(\mathcal{I}_{\gamma e}(\phi)+\overline{\mathcal{I}}_{\gamma e})$. We note the compensation that occurs when the two steps of Compton scattering, with a maximum at $\phi_{\trm{max}}=0$ and pair-creation with a maximum at $\phi_{\trm{max}}=\pi/2$ are combined, by the dot-dashed line in \figref{fig:DelRge}, which marks the corresponding maximum for the two-step process in the range $0.15\lesssim2\phi_{\trm{max}}/\pi\lesssim0.4$, with higher $\phi_{\trm{max}}$ values for lower $\chi$-values of the incoming electron. By combining the asymptotic limits from each sub-step, one can show:
\begin{equation}
\frac{2\phi_{\trm{max}}}{\pi}\sim
\begin{cases}
\frac{1}{2} & \chi_{1} \ll 1\\
0 & \chi_{1} \gg 1.
\end{cases}
\end{equation}

\subsection{Differential rate}
\begin{figure}[!h]
\centering\noindent
 \includegraphics[draft=false,width=6.25cm]{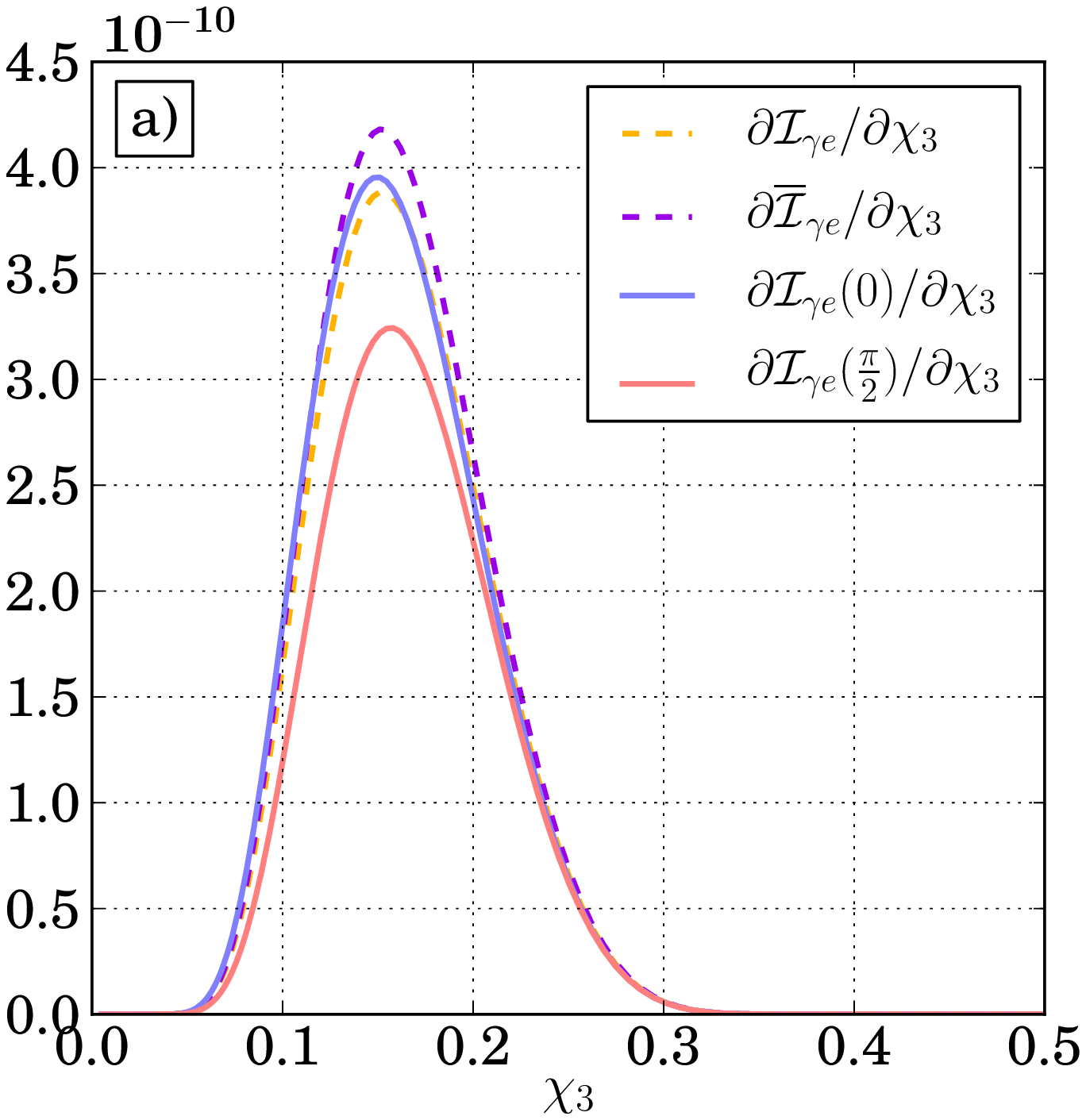} \\
 \includegraphics[draft=false,height=6.25cm]{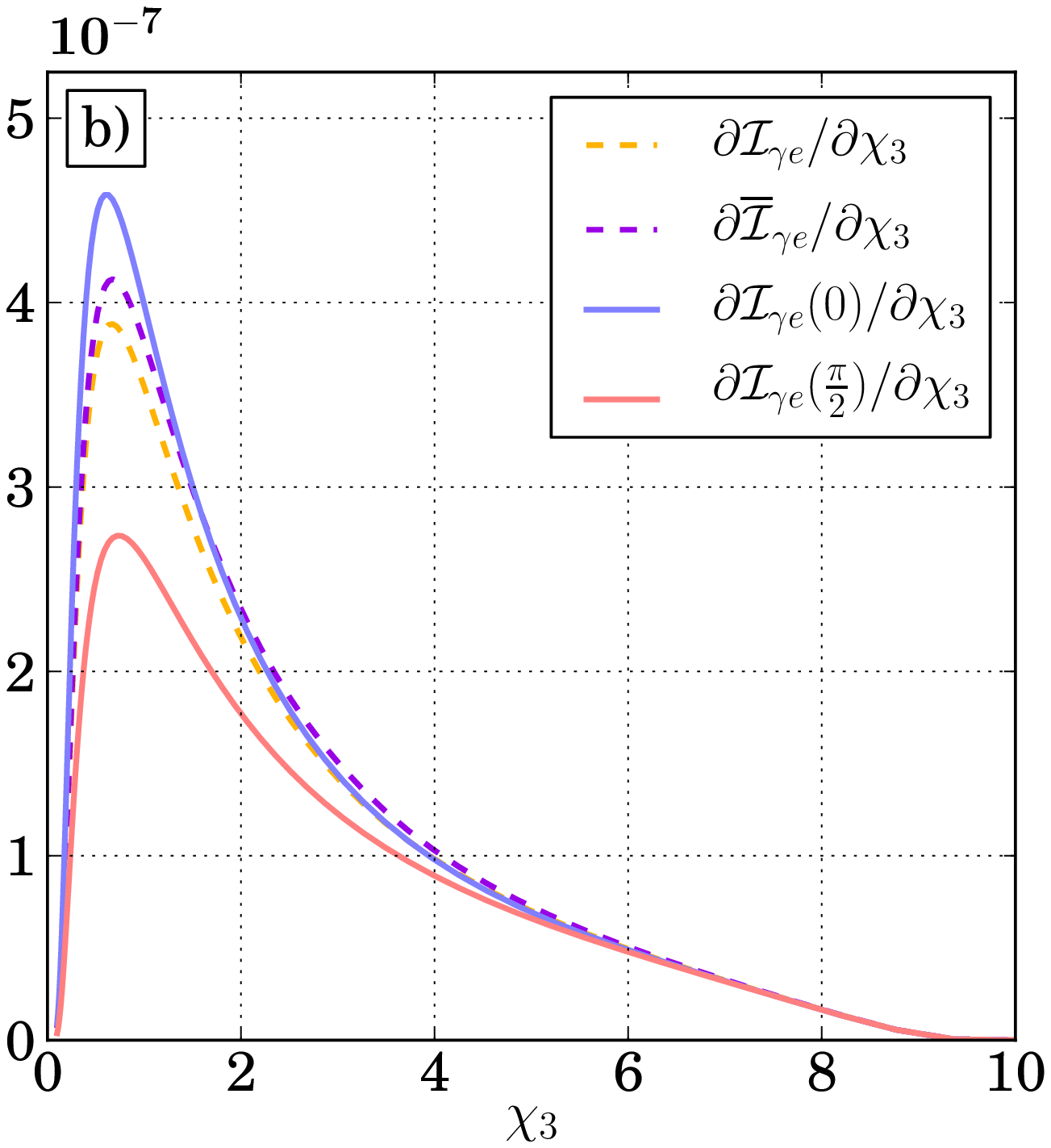}
\caption{(Color online). Plot of the differential rate for creating the next generation of pairs, given an incident
electron with $\chi_{1}=0.5$ (plot a)) and $\chi_{1}=10$ (plot b)). The solid lines for definite photon polarisation have only very slightly displaced maxima. }
\label{fig:diffRs}
\end{figure}

\begin{figure}[!h]
\centering\noindent
  \includegraphics[draft=false, width=7cm]{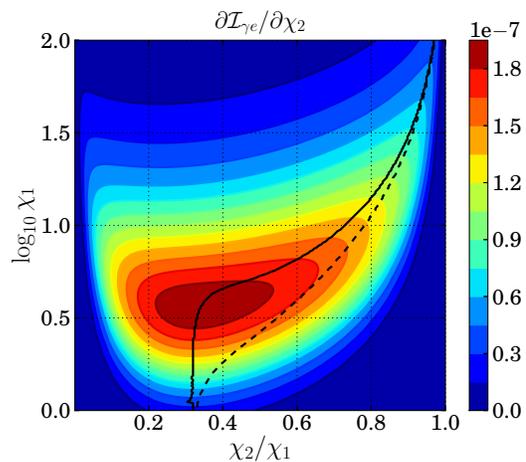}
\caption{(Color online). The differential rate of the electron step with respect to the scattered electron. The solid black line marks the most probable $\chi_{2}/\chi_{1}$ ratio for the scattered electron after emitting a real photon that decays into a pair and the dashed line marks the resulting value ($\chi_{2}$ becomes $\chi_{1}$ for the next generation) . The presence of a tail around $\chi_{2}=\chi_{1}$ hints at a cascading process.}\label{fig:diffR2}
\end{figure}

In \figref{fig:diffRs}, we plot the differential rate of the electron step  $\partial \mathcal{I}_{\gamma e}(\phi)/\partial \chi_{3}$ using unpolarised rates for each step ($\oI_{\gamma e}$), using $\phi=0,\pi/2$ and using the average over $\phi$, $\mathcal{I}_{\gamma e} = \langle \mathcal{I}_{\gamma e} (\phi) \rangle_{\phi}$.
The dynamics for pairs created with differently-polarised photons is very
similar, although the maxima are slightly displaced and there is
a slight asymmetry between $\partial \mathcal{I}_{\gamma e}/\partial\chi_{3}$ and
$\partial \oI_{\gamma e}/\partial\chi_{3}$ as compared to the fixed $\phi=0,\pi/2$
polarisations shown by lines crossing in the plots (the plots are identical for
$\chi_{3}\to \chi_{4}$ i.e. for electrons and positrons). It was noted that
for a higher incoming quantum nonlinearity parameter ($\chi_{1}$), the proportion of energy given to the
created pair becomes, on average, lower, as the distribution in $\chi_{3}\in
[0,\chi_{1}[$ becomes increasingly skewed towards the lower end. In fact, $\chi_{3}$ stays between
$0.5<\chi_{3}<0.75$ for $1<\chi_{1}<10^{3}$. This can be explained by noticing
that Compton scattering leading to pair creation is most probable around $\chi_{k}\approx 1$ and since
$\chi_{k}=\chi_{3}+\chi_{4}\approx O(2\chi_{3})$ (where $O(\cdot)$ corresponds to ``of the order of''), 
that the most probable value of $\chi_{3}$ for $\chi_{1}>1$ (hence allowing $\chi_{k}>1$) is around $\chi_{3}\approx
0.5$. One could conjecture the existence of the two types
of cascade mentioned in the introduction; a \emph{free-particle} and a
\emph{field-driven} cascade. For a high-$\chi$ incident fermion, it would seem
that each Compton-scattering-particle-creation event reduces the $\chi$-factor
only slightly. This is shown in \figref{fig:diffR2} where the differential
cross-section of $\mathcal{I}_{\gamma e}$ in $\chi_{2}$ (the scattered fermion) is
plotted, and the most probable ratio of $\chi_{2}/\chi_{1}$ is marked with the
solid black line. Since the rate $\mathcal{I}_{\gamma e}$ is expressed as a probability
per unit external-field phase, although large and small values of $\chi_{1}$ may
have the same value of $\mathcal{I}_{\gamma e}$, the probability that a
pair is created in a given duration in the lab-frame in a homogeneous field is
much higher for the higher value of $\chi$ as it traverses more external-field
phase than the lower-value $\chi$ particle. 
\figref{fig:diffR2} shows that if a $\chi_{1}\lesssim5$ fermion produces a pair, further acceleration from
the external field will be required before pair-creation becomes comparatively
probable again. This would represent a transition from the free-particle to the
field-driven cascade. We note that we have focused simply on the two-step
pair-creation process, but the two-photon Compton scattering process
\cite{seipt12,*mackenroth13} would most likely be more important to describe the
fermion dynamics, as single-photon Compton scattering is more probable than
pair-creation for all values of $\chi$ considered here.

\subsection{Total rate}
\begin{figure}[!h]
\centering\noindent
  \includegraphics[draft=false, width=6.25cm]{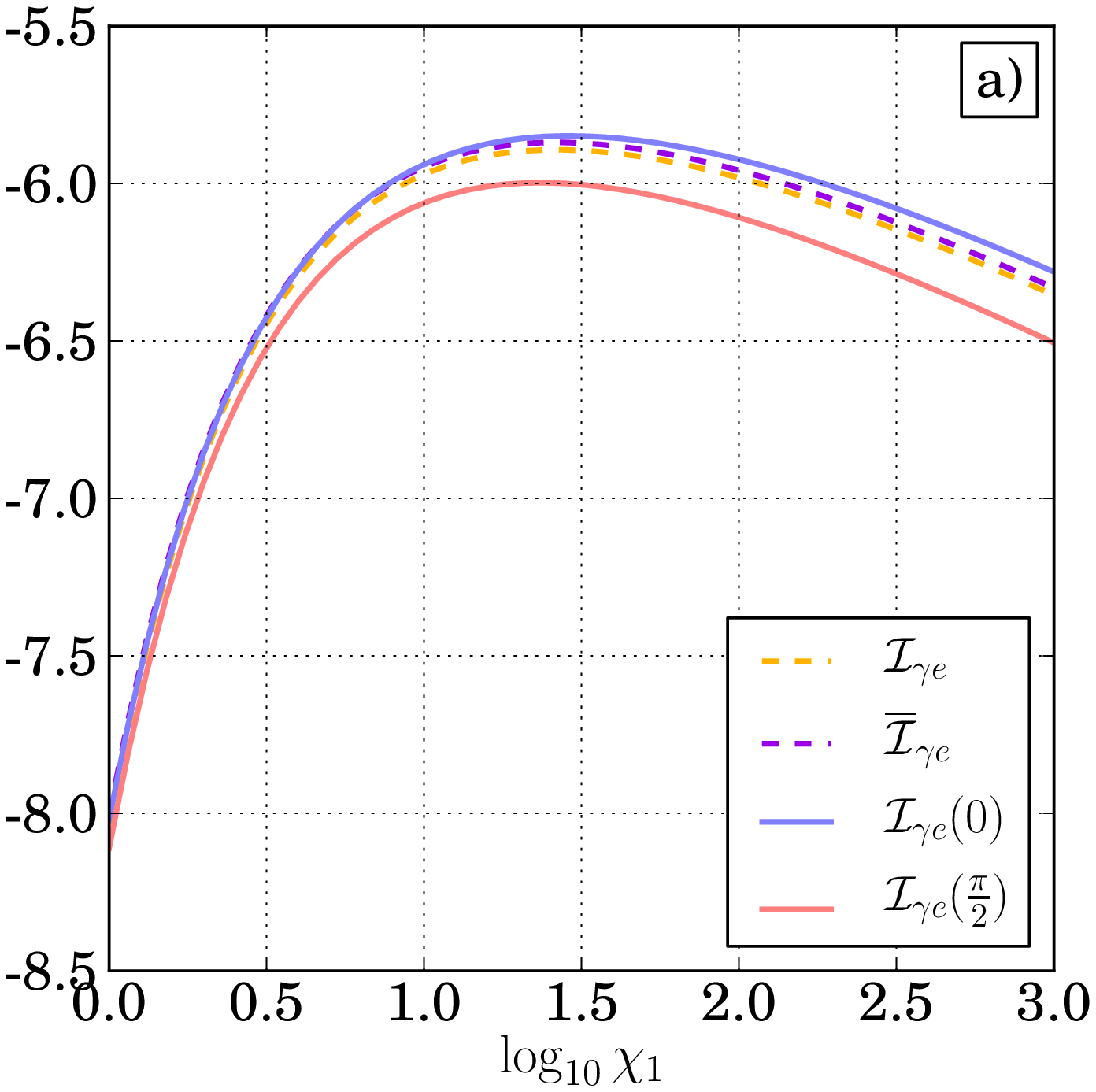}\\
 \includegraphics[draft=false,  width=6.25cm]{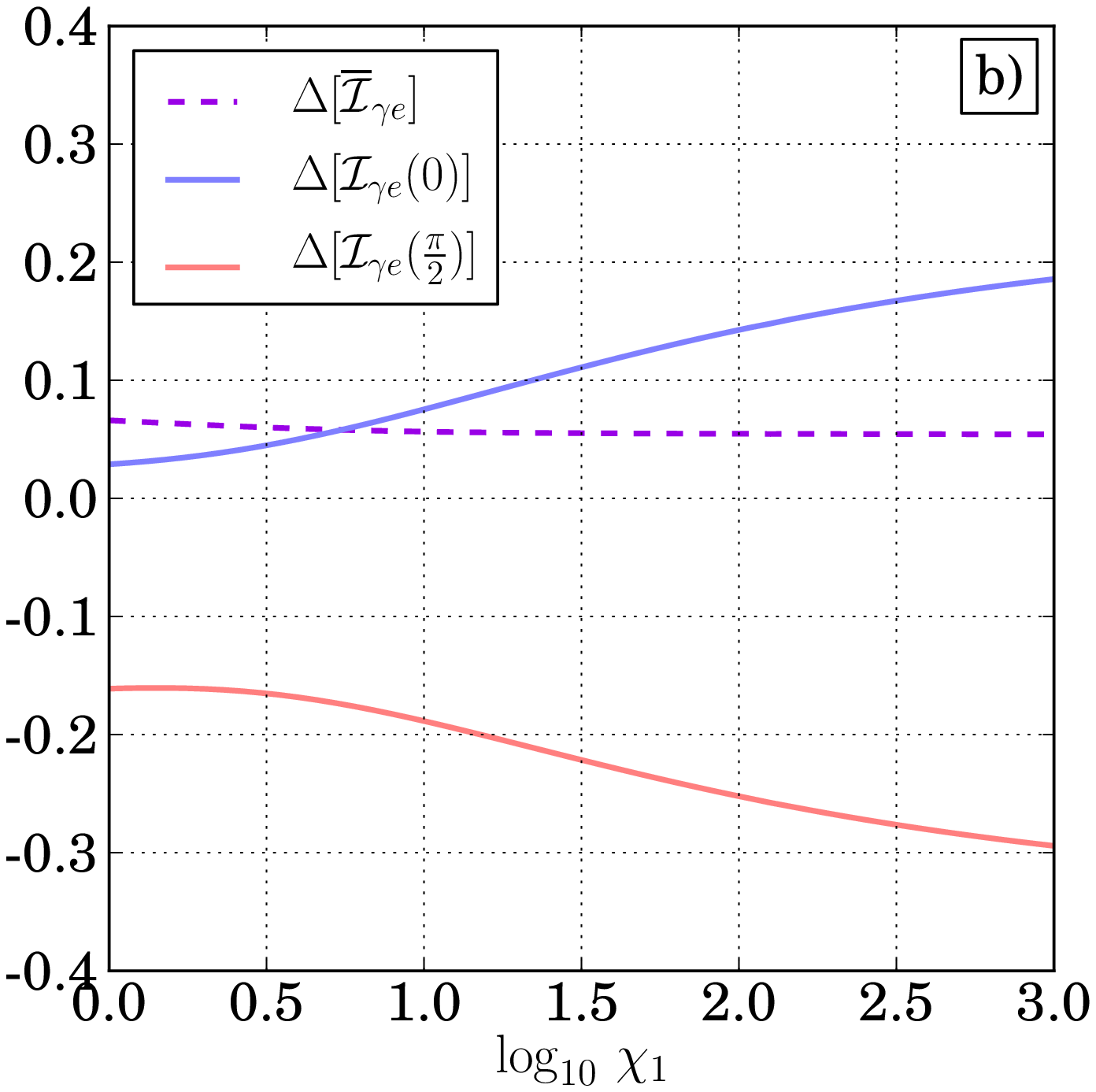}
\caption{(Color online). Plot of the total two-step process $\mathcal{I}_{\gamma e}$ (in a)) and the relative difference to using a polarised intermediate photon $\Delta[\mathcal{I}]\defto (\mathcal{I}-\mathcal{I}_{\gamma e})/\mathcal{I}_{\gamma e}$ (in b)).}\label{fig:DelPlot}
\end{figure}
By integrating under the curves in \figref{fig:diffRs}, we acquire the total fermion two-step rate, $\mathcal{I}_{\gamma e}(\phi)$, plotted in \figref{fig:DelPlot}. To deduce the overall difference that each polarisation makes, we plot the relative difference, $\Delta[\mathcal{I}]$ with respect to $\mathcal{I}_{\gamma e}$, $\Delta[\mathcal{I}]\defto (\mathcal{I}-\mathcal{I}_{\gamma e})/\mathcal{I}_{\gamma e}$, displayed in \figref{fig:DelPlot}. We notice on the one hand that the total rate for photons polarised in the $\phi=0,\pi/2$ direction differ by around $30\%$ for $\chi_{1}<10$, with the difference growing with $\chi_{1}$, unlike for tree-level rates. On the other hand, the difference between unpolarised $\oI_{\gamma e}$ and polarised $\mathcal{I}_{\gamma e}$ rates remains small at approximately $5\%$. This represents a smoothing out of the larger relative differences found for the individual processes in \figrefs{fig:Rgs}{fig:Res} suggesting the polarisation correlation of the form $\langle\int dv (\partial \mathcal{I}_{\gamma}(\phi,v)/\partial v) \mathcal{I}_{e}(\phi,v) \rangle_{\phi} - \int dv \langle (\partial \mathcal{I}_{\gamma}(\phi,v)/\partial v) \rangle_{\phi}\langle \mathcal{I}_{e}(\phi,v) \rangle_{\phi}$ is weak and the approximation of using unpolarised rates in simulations is valid. In order to investigate the effect of polarisation when a greater variety of chains of processes occur, we turn to simulation.

\section{Cascade simulation} \label{sxn:CascSim}
We wish to investigate the cumulative effect of photon polarisation when Compton-scattering and pair-creation processes form a cascade. To this end, we employ simulation methods developed in \cite{ruhl10, elkina11}, which integrate over these lowest-order rates to approximate chains of events. Simulation has the veritable advantage that many possible chains of real processes are considered, for example that several Compton-scattering steps can occur before a pair-creation step, which for some values of $\chi_{1}$ have the potential to expose the polarisation behaviour. Moreover, although we have analytically investigated the idealised background of a constant crossed field, in a simulation, one can employ the so-called ``locally-constant field approximation,'' in which the constant crossed field rates are integrated over the phase of a more complicated field. In this case, $\chi \to \chi[E(\varphi)]$, where $E$ is the electric field amplitude with a more complicated structure and $\varphi$ is its phase. We take the range of validity of such an approximation to be the same as the validity of the constant crossed field expression, already discussed in the paragraph below \eqnref{eqn:paramdef}. However, the higher the variety of chains of events, the more challenging it is to directly compare with theory.
\newline

\begin{figure}[!h]
\centering\noindent
 \includegraphics[draft=false, width=8.6cm]{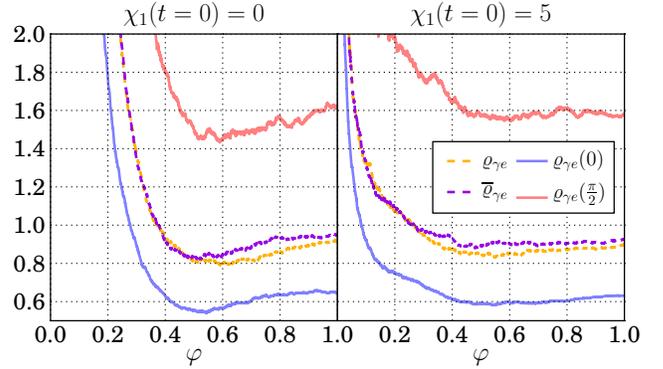}
\caption{(Color online). A plot of the ratio of total fermions created to hard ($\chi_{k}\geq 1$) photons $\varrho_{\gamma e}$. Although pairs are created much earlier in the $\chi_{1}(t=0)=5$ than in the $\chi_{1}(t=0)=0$ case, after just $\varphi=1$, the ratios $\varrho_{\gamma e}$ is of the order of unity. }\label{fig:Dups}
\end{figure}

In order to incorporate the polarised cross-sections, we rewrite the rate equations in the simulation model (Eqs. (2,3) in \cite{elkina11}) in terms of rates per unit energy $\mathcal{E}$:
\bea
\frac{dW_{\gamma}}{d\mathcal{E}_{\gamma}} &=& \frac{-\alpha m^{2}}{\mathcal{E}_{e}^{2}}\left\{\Ai_{1}(x) +\left(\frac{g(\phi)}{x} + \chi_{k} \sqrt{x}\right) \Ai'(x) \right\},\nonumber\\
\frac{dW_{e}}{d\mathcal{E}_{e}} &=& \frac{\alpha m^{2}}{\mathcal{E}_{\gamma}^{2}}\left\{\Ai_{1}(x) +\left(\frac{g(\phi)}{x} - \chi_{k}\sqrt{x} \right)\Ai'(x) \right\},\nonumber\\
\eea
where $g(\phi)=2\cos^{2}\phi + 1$, $x=\mu^{2/3}$. As an example scenario, we simulate the presence of $10^{3}$ initial electrons in a rotating electric field $\mbf{E}(t) = (E_{0}\cos \varphi, E_{0} \sin \varphi, 0)$, where $\varphi=\varkappa^{0} t$, $\varkappa^{0}=1~\trm{eV}$ is the angular frequency and $\xi=10^{4}$ ($\chi_{E} \approx 0.02$) (the strong-field QED effects in this field are to a good approximation equivalent to those of a constant crossed field background, see also \cite{elkina11}), for two cases: electrons initially at rest ($\chi_{1}(t=0) = 0$) and initially counter-propagating with $\chi_{1}=5$ against the field. Each simulation is run until $\varphi=\vkap^{0} t=1$. For each scenario, the four different cases are simulated in which: i) the parameter $\phi$ is randomly selected from a uniform distribution $\phi \in [0, \pi/2[$ (the physical case) with quantities $N$ denoted $N_{\gamma e}$; ii) unpolarised rates are used denoted by $\overline{N}_{\gamma e}$; iii) 
$\phi = 0$, denoted by $N_{\gamma e}(0)$; iv) $\phi=\pi/2$, denoted by $N_{\gamma e}(\pi/2)$. 
\begin{figure}[!h]
\centering\noindent
 \includegraphics[draft=false, width=6.25cm]{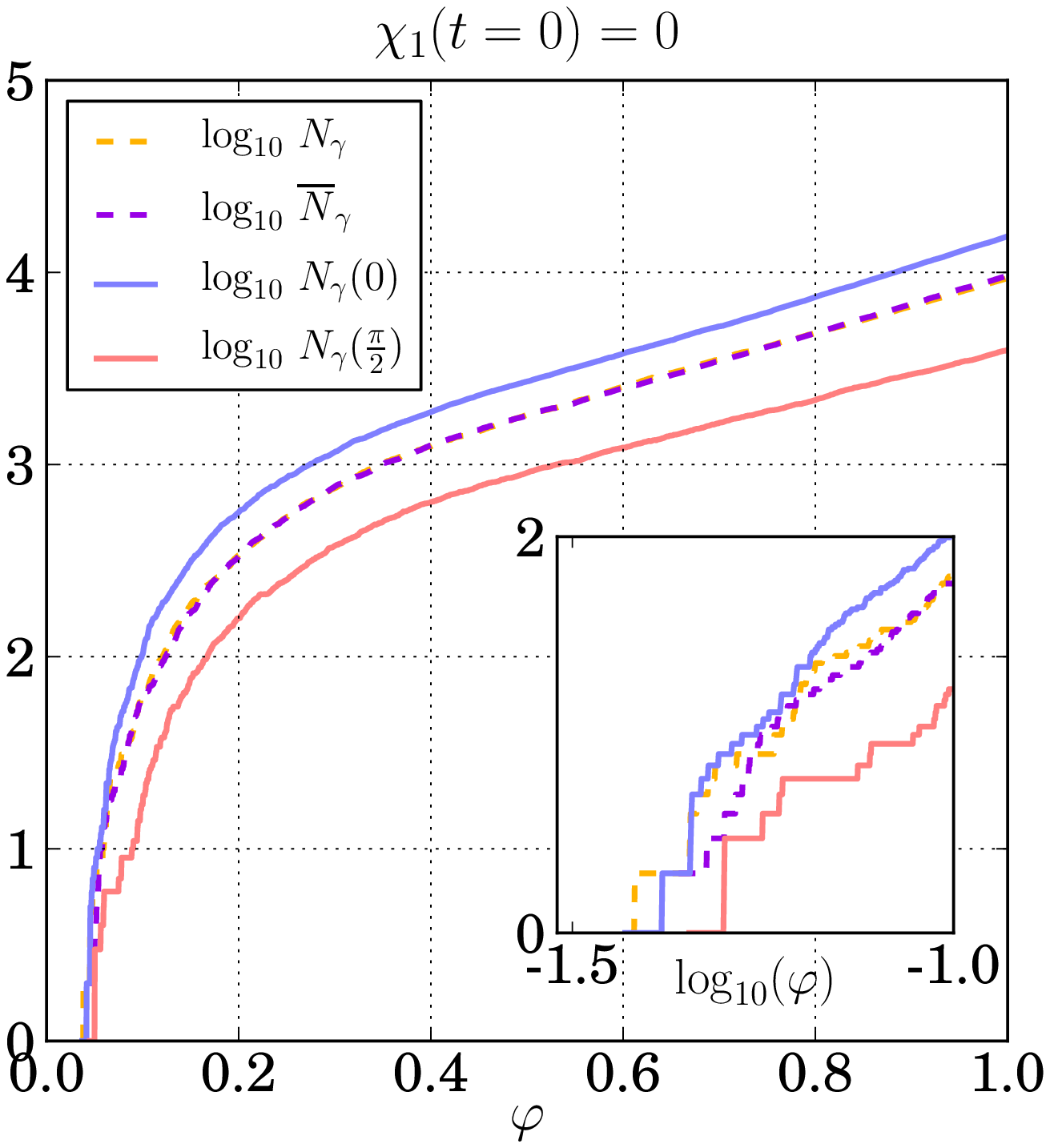}  \\
\includegraphics[draft=false, width=6.25cm]{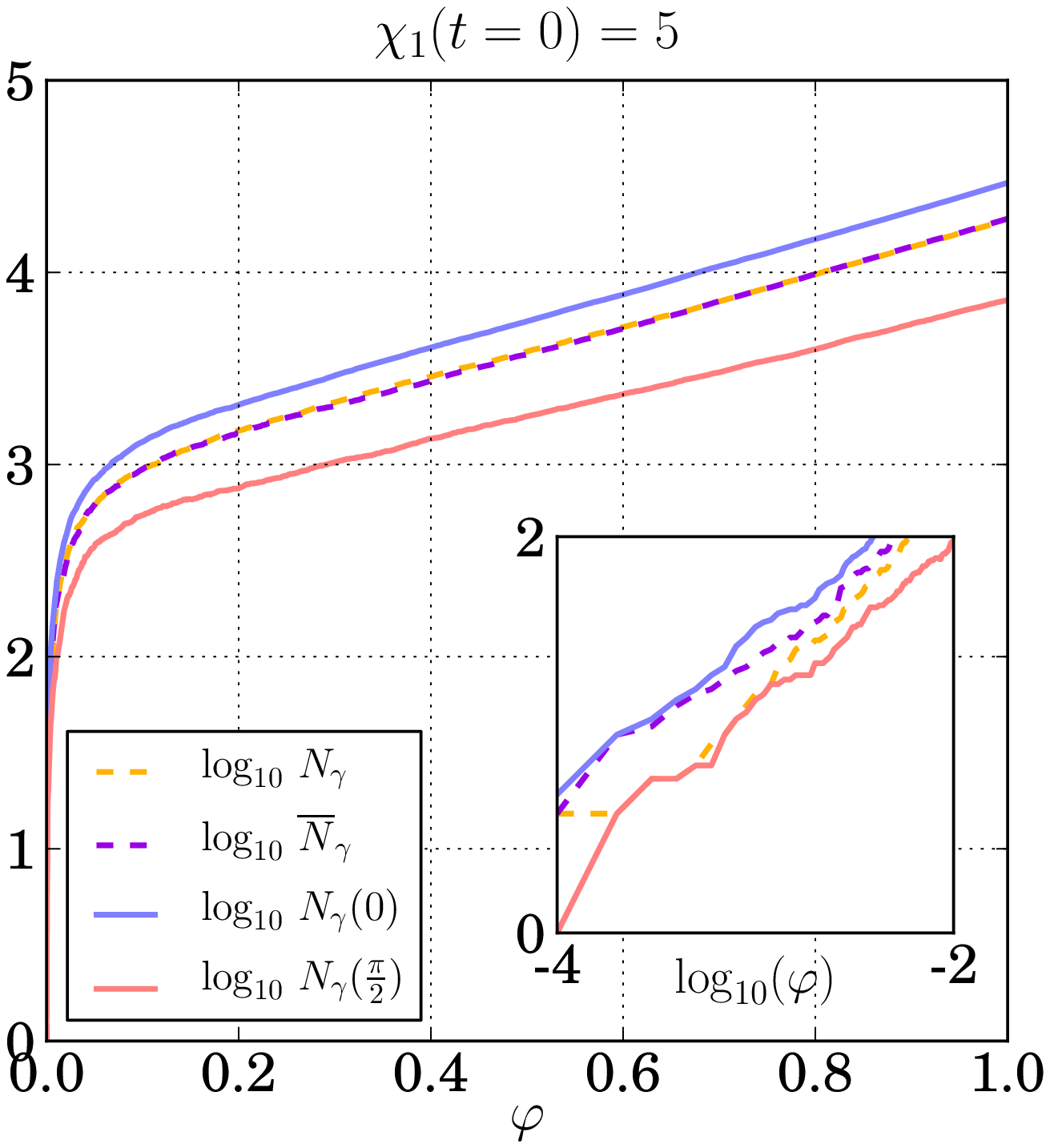}
\caption{(Color online). A plot of the logarithm of the number of hard photons ($\chi_{k}\geq1$) when $10^{3}$ electrons initially with $\chi_{1}=0, 5$ respectively, interact with a rotating electric field of frequency $\varkappa^{0}=1~\trm{eV}$, $\xi=10^{4}$. In the inset is a log-log plot of the initial stages of the cascade.}\label{fig:Ng_chi1s}
\end{figure}
\begin{figure}[!h]
\centering\noindent
  \includegraphics[draft=false, width=6.25cm]{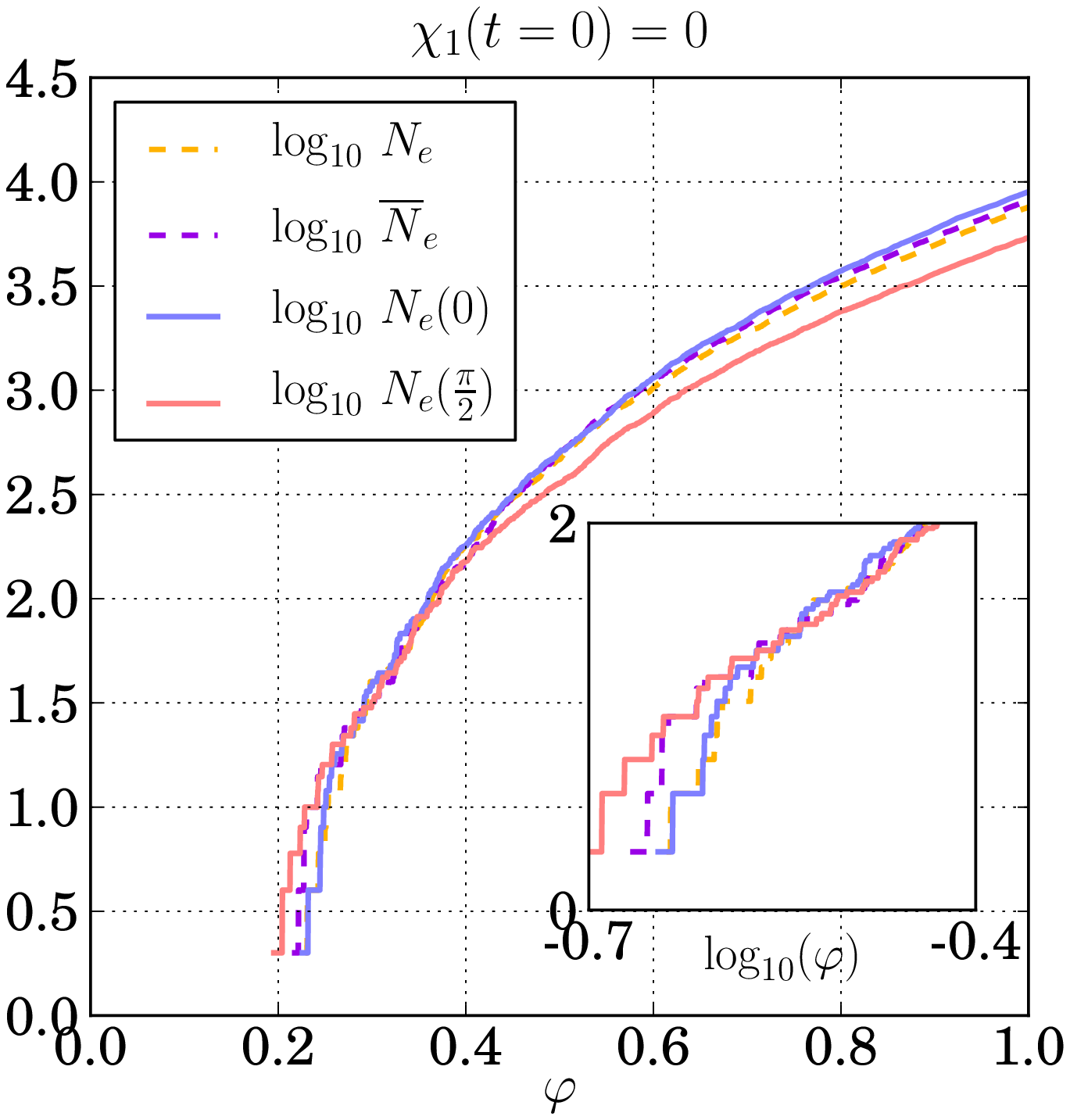} \\
\includegraphics[draft=false, width=6.25cm]{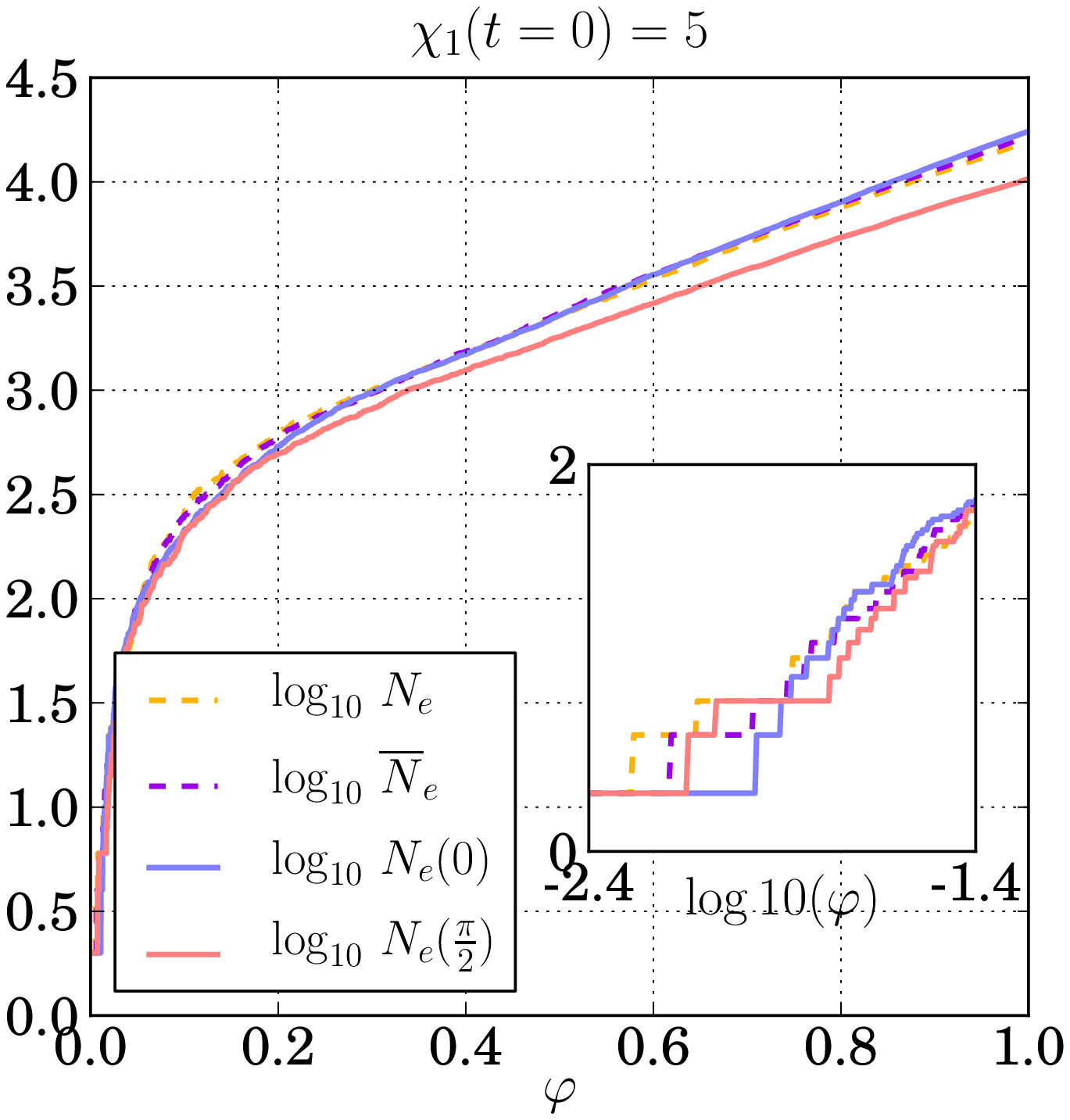}
\caption{(Color online).  A plot of the logarithm of the number of fermions created in the test simulations, $\xi=10^{4}$, $\varkappa^{0}=1~\trm{eV}$ for $\chi_{1}(t=0)=0$ and $\chi_{1}(t=0)=5$. When production begins, there is a jump in $N_{e}$ due to charge conservation ($N_{e}$ is an even number).}\label{fig:Ne_chi1s}
\end{figure}
The number of photons with $\chi_{k} > 1$ and fermions with $p^{0} > 20m$ ($\chi>0.4(1-\cos\theta)$, where $\theta$ is the angle between $\mbf{p}$ and $\pmb{\varkappa}$) generated with the above parameters are plotted in \figrefs{fig:Ng_chi1s}{fig:Ne_chi1s} respectively. The straight part of the plots represents an equilibrium between momentum change due to QED processes and acceleration by the field. In \figref{fig:Dups}, we plot the ratio of photons to fermions, which is found to be of the order of unity, although the number of fermions created by $\varkappa^{0}t=1$ shows that, depending on photon polarisation, the average number of generations in the cascade is between $1.8$ and $3.3$ (using $N_{e}=2(2^{n}-1)N_{\gamma}(t=0)$ as the number of fermions created after $n$ generations).
\subsection{Photon sector}
\subsubsection{Polarisation behaviour}
\begin{figure}[!h]
\centering\noindent
  \includegraphics[draft=false,width=8.6cm]{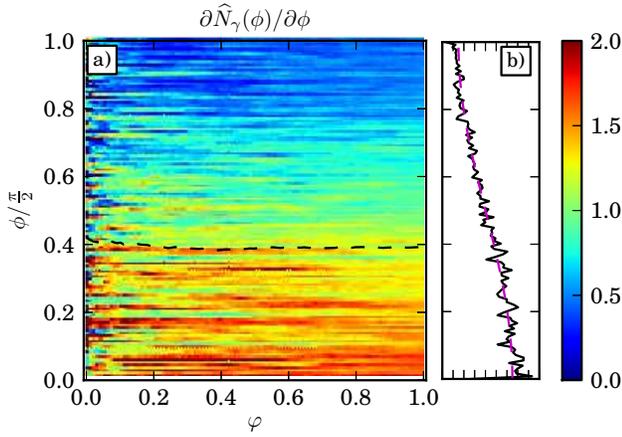}
\caption{(Color online). In plot a) is the normalised distribution of photons against polarisation angle (for $\chi_{1}(t=0)=5$) with the dashed line corresponding to the average polarisation. After an initial transient region, the distribution becomes smooth but retains its shape. Plot b) compares $N_{\gamma}$ at $\varphi=1$ against the normalised rate for Compton scattering (dashed line), averaged over $\chi_{1}\in[1,10]$.}\label{fig:polRes}
\end{figure}

Before describing the evolution of photon and fermion number for different polarisations, it is important to understand how the distribution of photon polarisations evolves. For the case when electrons have an initial $\chi_{1}=5$, the distribution of polarisations normalised so that $\int_{0}^{\pi/2}\!d\phi\,\partial \widehat{N}_{\gamma}(\phi)/\partial\phi=1$ is plotted in \figrefa{fig:polRes}. After an initial transient period, the occupation of polarisation angles smooths out and a distribution forms, which appears constant in time. Alongside this in \figrefb{fig:polRes}, $\partial \widehat{N}_{\gamma}(\phi)/\partial\phi$ is plotted at $\varphi=1$ and shows excellent agreement with the plot of $R_{\gamma}(\phi)$ averaged over $\chi_{1}\in[1,10]$ (typical values for the simulation). By using polarised Compton scattering and pair-creation rates, there is a correlation between these two steps, but only in this order. In the simulation, the next generation of fermions then Compton scatter without any influence from previous steps. The photon polarisations are then distributed as they would be due to single, incoherent, Compton scattering events. The potential smoothing of this distribution due to pair creation appears not to take place, which could be understood when one realises that hard photons are produced more easily than pairs (as there were no seed photons, the number of photons is necessarily greater or equal to the number of pairs generated). In reality, one might expect that the polarisation distribution would evolve with the plasma. What is missing from this model for this to take place is fermion spin correlation between the stages of pair-creation and Compton-scattering, which would carry the influence of photon polarisation over successive generations. Another approximation used that combining the rate of single events is equal to the rate of the chain of these events has been recently supported by calculations in \cite{king13b}. 

\subsubsection{Photon population behaviour}
\begin{figure}[!h]
\centering\noindent
\includegraphics[draft=false,width=4.25cm]{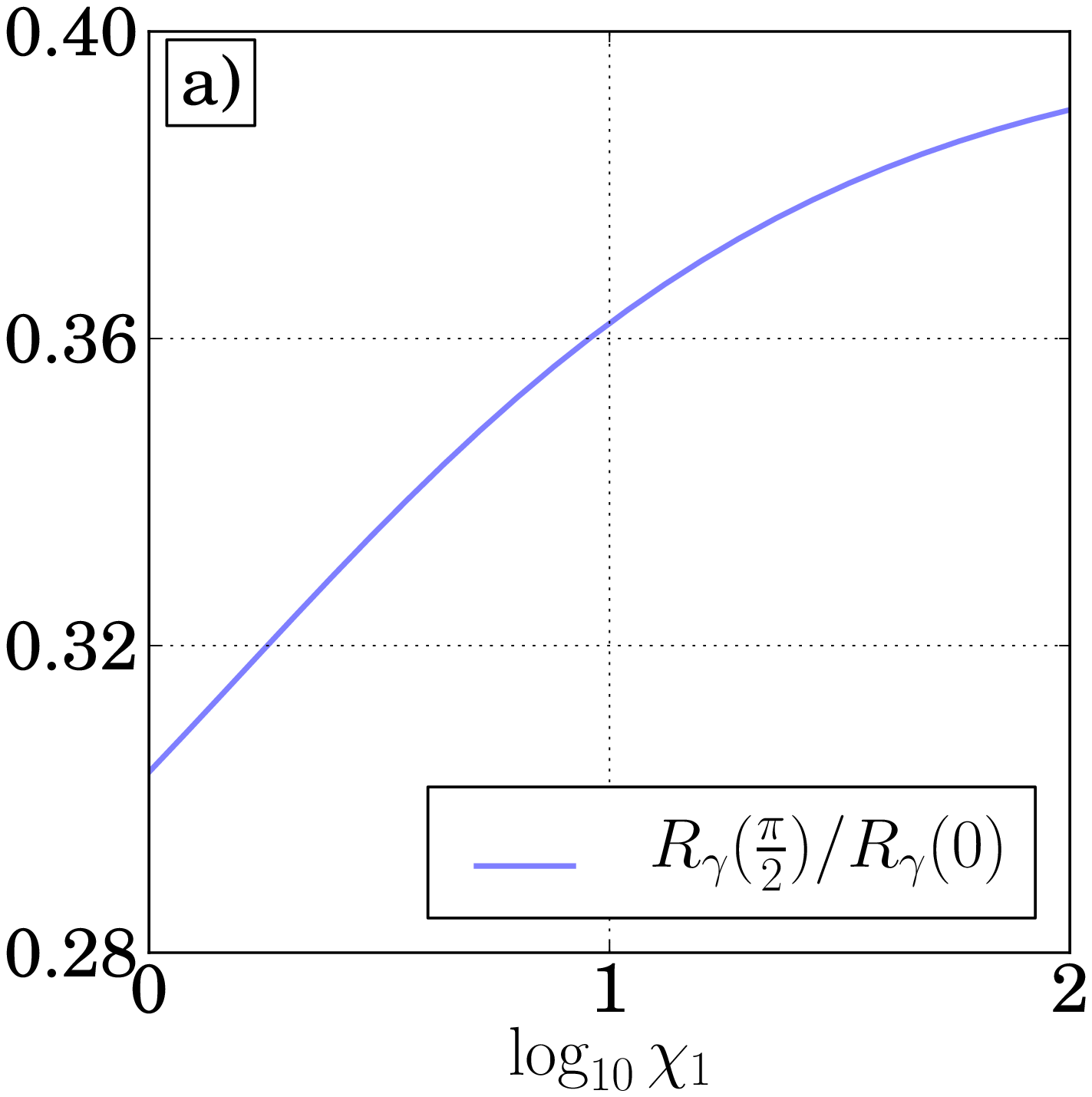}
\includegraphics[draft=false, width=4.25cm]{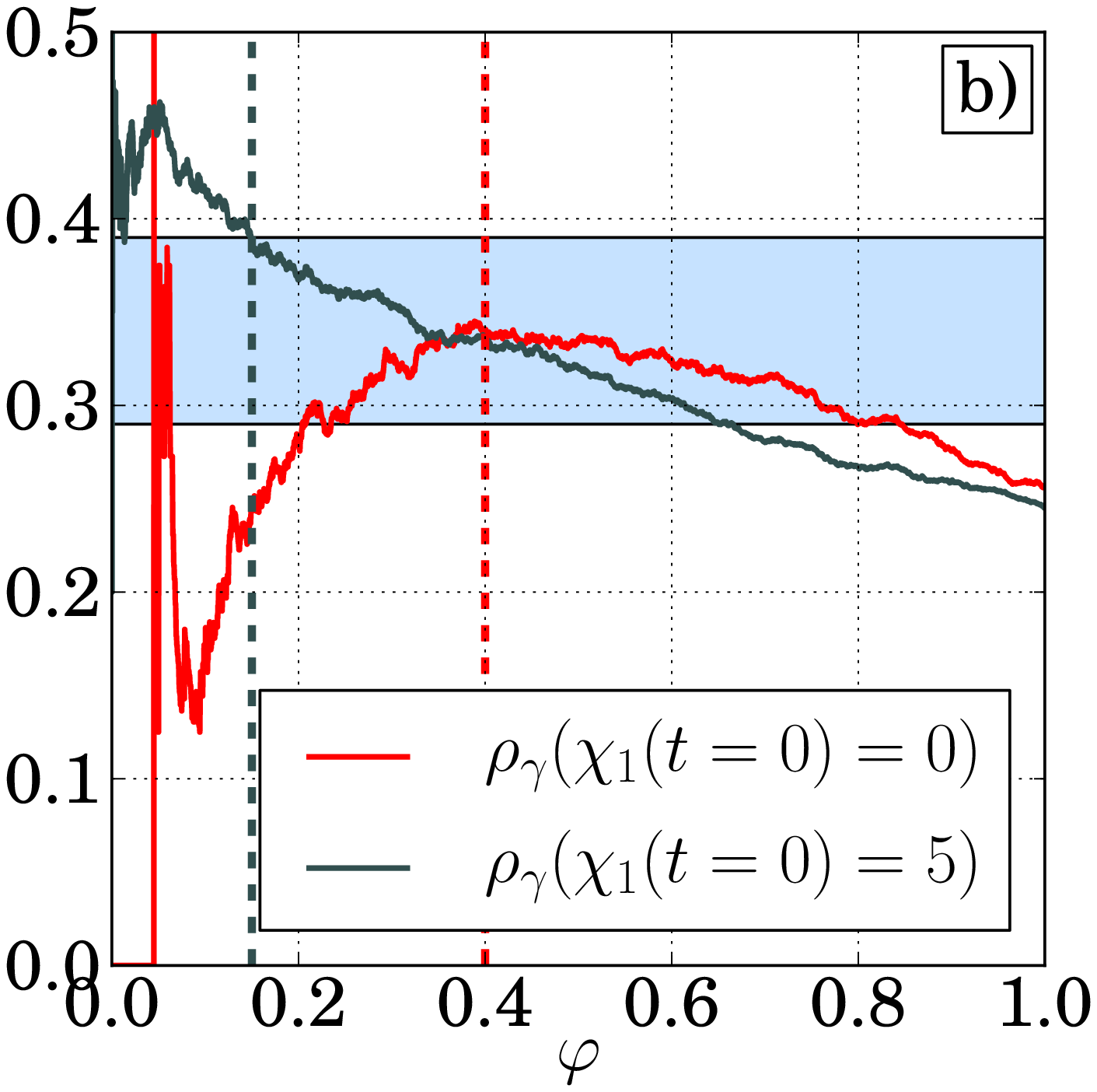}
\caption{(Color online). A comparison of the ratio of Compton-scattering rates for the polarisations $\phi=\pi/2, 0$ for a range of $\chi_{1}$ relevant to the simulation a), with the ratio of numbers of photons generated in the simulation $\rho_{\gamma}$, b). The bounds of the shaded region in b) are given by the left-hand plot, and the vertical dashed lines denote when the ``equilibrium'' phase of constant exponential growth is entered.}\label{fig:NgdiffR2}
\end{figure}

In a plot of the numbers of photons generated in \figref{fig:Ng_chi1s}, we make the following observations: i) the production of photons when $\phi=0$ is set is considerably more probable than when $\phi=\pi/2$ is chosen; ii) the difference in the number photons generated using unpolarised and polarised rates is very small (this was also reflected in the frequency spectra) and iii) the time required for photon-production to begin when $\chi_{1}(t=0)=5$ was more than two orders of magnitude larger for the $\chi_{1}(t=0)=0$ case. For the first point, the ratio of $\phi=\pi/2$ to $\phi=0$ photons can be verified by calculating the ratio of the rates of photon production in each of the two cases. As this is a comparison with one scattering event, it should be mainly useful when the number of Compton scattering events per fermion is low, as when many generations have been created, the biasing of events due to forcing $\phi$ to take a specific value should become evident. In \figref{fig:NgdiffR2} we verify the ratio of numbers of photons $\rho_{\gamma} =N_{\gamma}(\pi/2)/N_{\gamma}(0)$ by comparing the theoretical rate (\figrefa{fig:NgdiffR2}) with the yield from simulation (\figrefb{fig:NgdiffR2}), where the single Compton-scattering formula shows good agreement.
\newline

For the second point, as commented on in previous sections, taking the unpolarised rate is equivalent to taking the average rate. So the observation that taking a random polarisation of photon or taking the unpolarised rate makes little difference is simply indicative that after thousands of Compton-scattering events, using the average rate for each event is a good approximation. Furthermore, this is supported by noting that $N_{\gamma}$ and $\overline{N}_{\gamma}$ are at the average of the positions of $N_{\gamma}(0)$ and $N_{\gamma}(\pi/2)$ just as for the predicted average rate. The final point about the time of onset of Compton scattering being larger for $\chi_{1}(t=0)=0$ is also intuitive. As $P_{\gamma} = (m\chi_{E}L_{\varphi}/\varkappa^{0})\mathcal{I}_{\gamma}$, where $L_{\varphi}$ is the phase length, the expected 
number of photons generated in the first timestep $L_{\varphi}=10^{-4}$ for $\chi_{1}=5$, $\chi_{E}=0.02$ is $\approx 4$. It then follows that in the simulation,  Compton scattering can proceed immediately, which is reflected in \figref{fig:Ng_chi1s}. For $\chi_{1}(t=0)=0$, the seed electrons must first be accelerated before they can produce photons. In a rotating electric field, one can show that the phase required for one of these electrons to reach $\chi_{1}=1$ is $\varkappa^{0}t_{\ast} = \sqrt{\varkappa^{0}/m}/\chi_{E}$ (see \cite{fedotov10}). This corresponds to $t_{\ast} \approx 0.07$, which is comparable with $t_{\ast} \approx 0.05$ from the simulation in \figref{fig:Ng_chi1s}.


\subsection{Fermion sector}
\begin{figure}[!h]
\centering\noindent
  \includegraphics[draft=false,width=4.25cm]{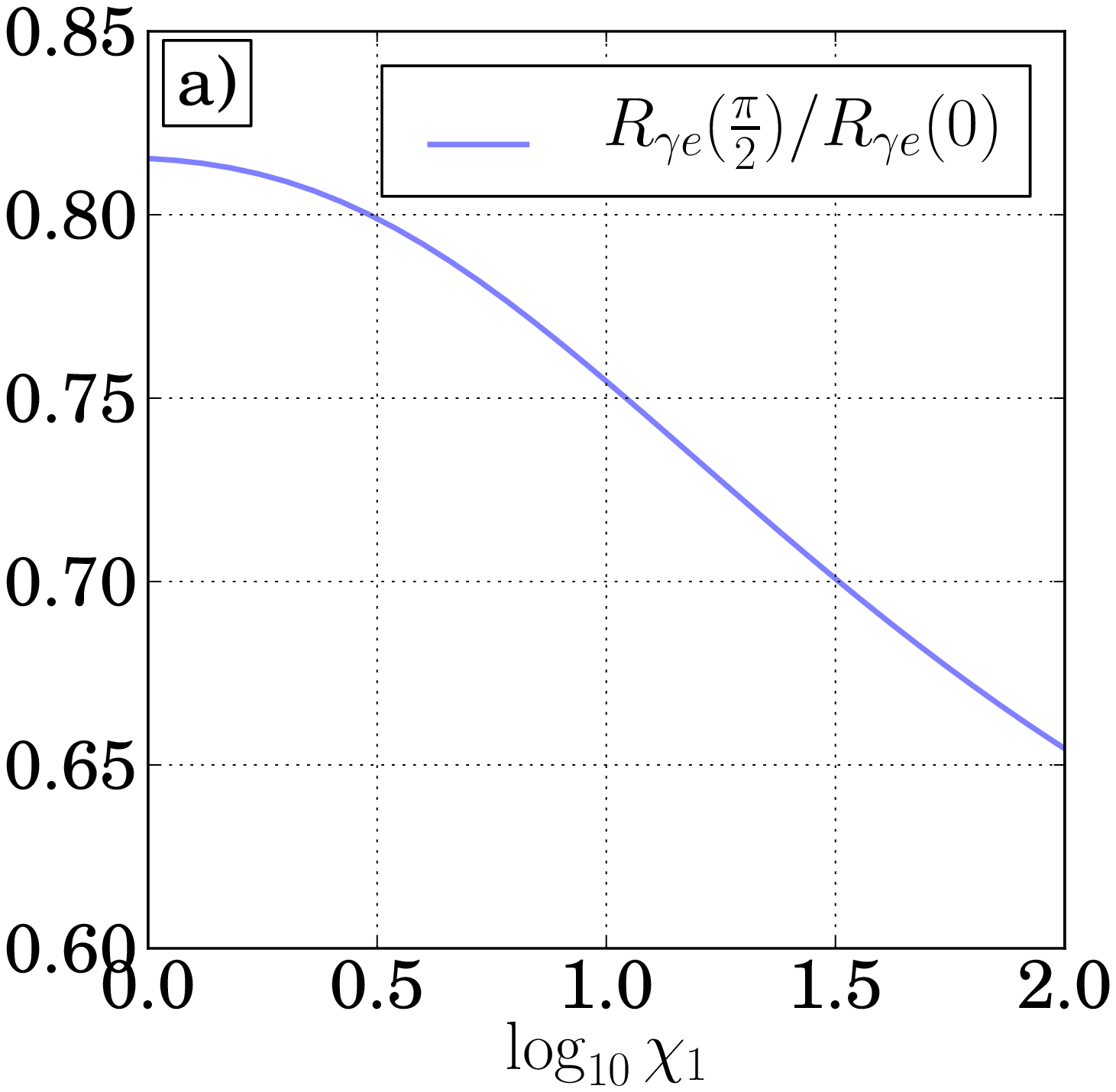}
 \includegraphics[draft=false,width=4.25cm]{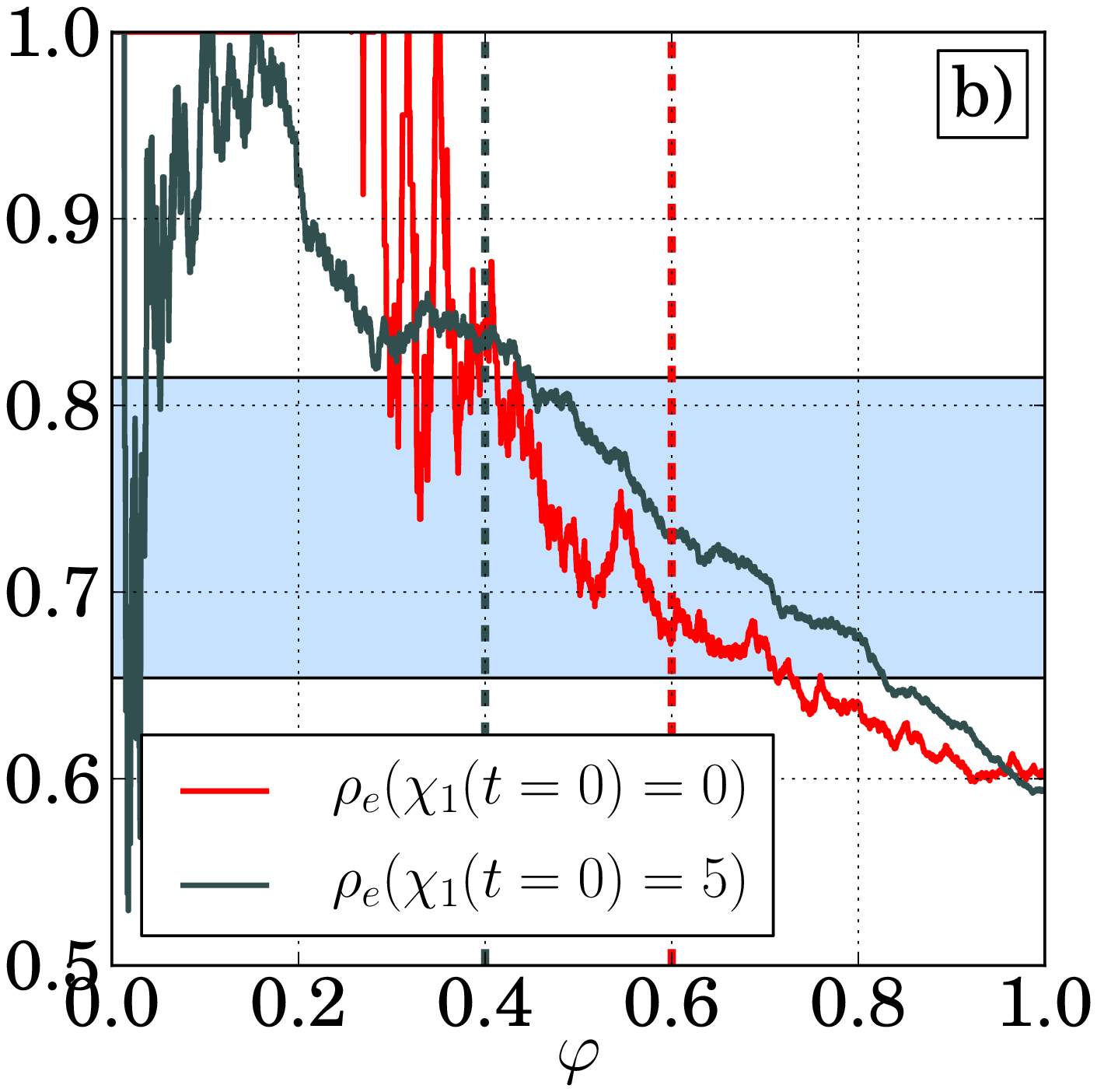}
\caption{(Color online). A comparison of the ratio of fermion-seeded pair-creation rates for the polarisations $\phi=\pi/2, 0$ and a range of $\chi_{1}$ found in the simulation (plot a)), with the ratio of numbers of photons generated in the simulation $\rho_{e}$ (plot b)). The bounds of the shaded region in b) are given by the theoretical prediction in a), and the vertical dashed lines in b) denote when the ``equilibrium'' phase of constant exponential growth is entered.}\label{fig:NfdiffR2}
\end{figure}

For the plot of the number of created fermions in \figref{fig:Ne_chi1s}, we 
make the following observations: i) although the photon-seeded pair-production
rate satisfies $R_{e}(\pi/2) > R_{e}(0)$, we notice $N_{e}(\pi/2)<N_{e}(0)$; ii) the
difference in the numbers of created pairs using the unpolarised and polarised
rate is again small although they are no longer in the middle of the maximum and
minimum curves as was the case in the photon sector iii) the time taken for
pair-creation to ensue when the electrons were initially at rest is orders of
magnitude larger than for $\chi_{1}(t=0)=5$. The first point can be explained by
noting that any pairs created must have gone through a process of $N$-fold
Compton scattering ($N\geq1$) followed by photon-seeded pair creation. Since
pair-production from any low-energy photons created by Compton scattering is
exponentially suppressed $R_{e}(\chi_{k}\ll1) \sim
\chi_{k}\mbox{e}^{-8/3\chi_{k}}/k^{0}$ \cite{nikishov64, kibble64}, we can
surmise that only $\chi_{k}\geq 1$ photons are relevant on the simulated time scales for 
pair-creation (in the simulations, a useful approximation was implemented that only $\chi_{k}\geq 1$ photons
were permitted to create pairs). We approximate the ratio of
pairs generated from photons polarised with $\phi=0,\pi/2$: $\rho_{e} =
N_{e}(\pi/2)/N_{e,1}(0)$ using the two-step probability in \eqnref{eqn:twostep}. This
approximation is plotted in \figref{fig:NfdiffR2} and predicts the correct range
of values for around the beginning of the equilibrium period. For Compton-scattering 
this is a much smaller ratio, and so when larger numbers of photons are 
Compton-scattered between pair-creation events, this ratio will be reduced compared 
to the theoretical prediction of single a Compton scattering before pair-creation, as observed 
in the numerics. For point ii), using again the two-step probability, we note that in the plot of the total rate
(\figref{fig:DelPlot}), $\mathcal{I}_{\gamma e}$ and $\oI_{\gamma e}$ are very close to
and even larger than for the maximum polarisation $\mathcal{I}_{\gamma e}(0)$ for
$\chi_{1}< 5$. This is reflected in both plots in
\figref{fig:Ne_chi1s}, where the case $\chi_{1}(t=0)=5$ clearly shows this
behaviour for times before equilibrium (in equilibrium, typically $\chi_{1}>5$ is quite possible \cite{elkina11}). 

\subsection{Discussion}
\begin{figure}[!h]
\centering\noindent
  \includegraphics[draft=false,width=4.25cm]{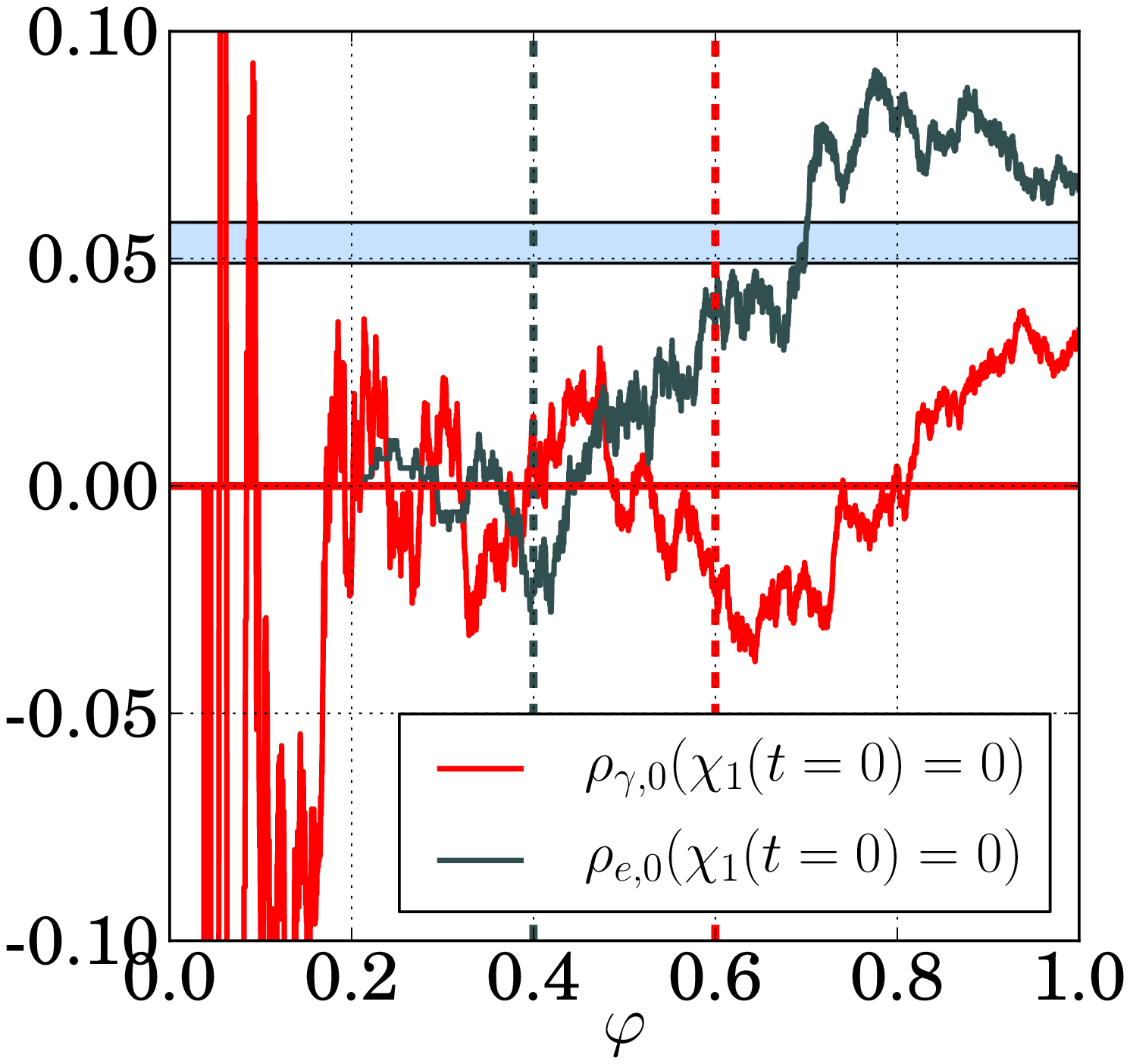}
\includegraphics[draft=false,width=4.25cm]{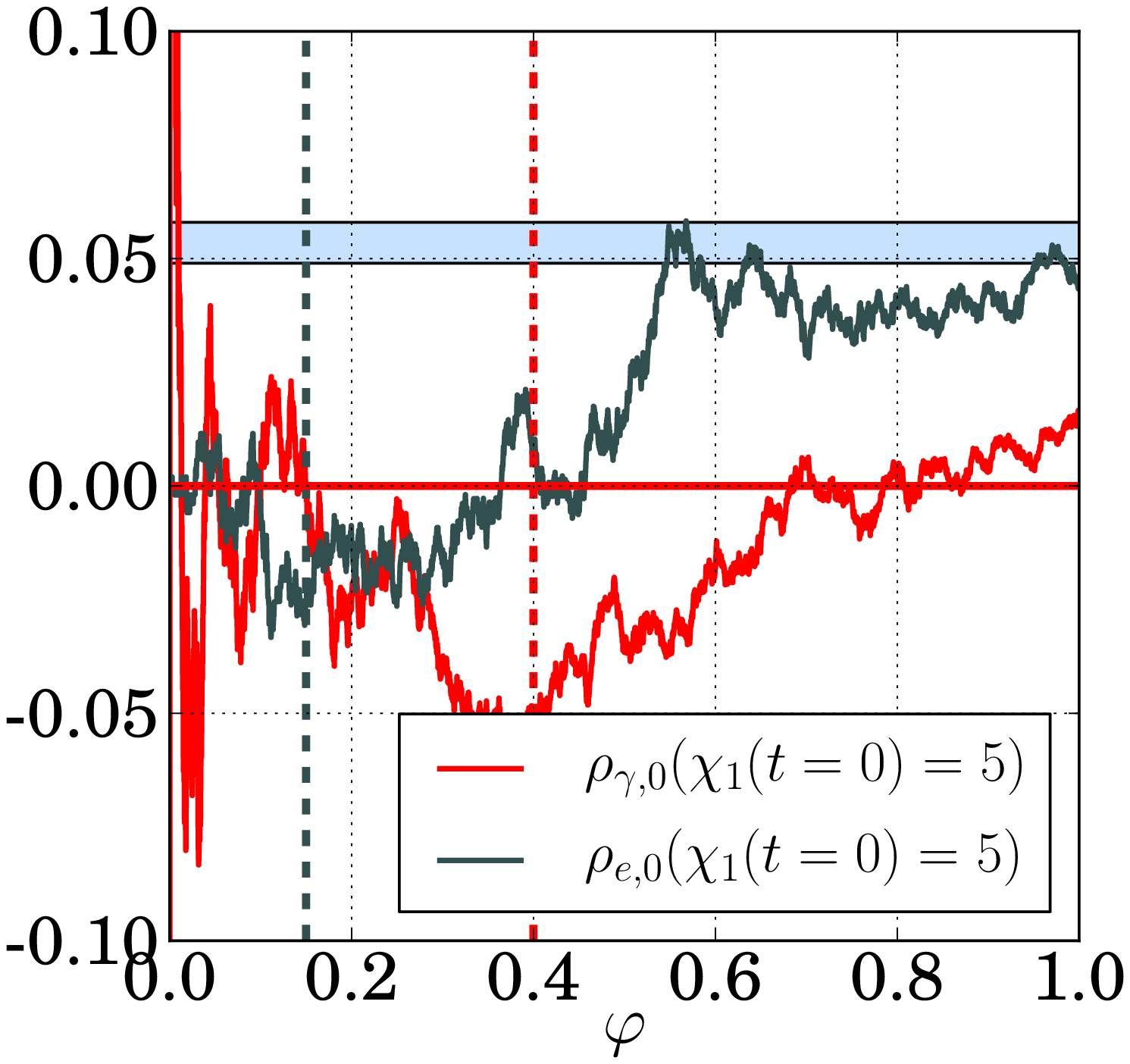}
\caption{(Color online). In the plot are the relative differences between the number of photons and pairs produced for the polarised and unpolarised case ($\rho_{\cdot,0} = N_{\cdot}-\overline{N}_{\cdot}/N_{\cdot}$). The solid red horizontal line and grey-blue zones are the predictions using arguments based on the rate for the respective curves.}\label{fig:RelPols}
\end{figure}

The results of numerical simulation in the current section and theoretical
analysis of \sxnreft{sxn:Comp}{sxn:CascTh} are in broad
agreement. Moreover, the numbers of particles created when
polarised tree-level rates were used agreed to within $5\%$ of when unpolarised
tree-level rates were used. This is verified in \figref{fig:RelPols} where the theoretical prediction from \figrefb{fig:DelPlot} for pair creation and the prediction of zero difference for Compton scattering are compared with the relative yields from simulation $\rho_{\cdot,0} =
N_{\cdot}-\overline{N}_{\cdot}/N_{\cdot}$ for photons ($\rho_{\gamma,0}$) and pairs
($\rho_{e,0}$) which, taking into account statistical
fluctuations, are around the same order of magnitude.  
One can surmise that the large number of seeds in the simulation aided this
polarisation-averaging effect, which would be reduced when the number of events
is small (quantum stochasticity in Compton scattering has recently been explored 
in \cite{neitz13}). If photon polarisation could be controlled 
by its environment, it was shown that the number of photons generated with polarisation $\phi=0$ could 
be much larger than with polarisation $\phi=\pi/2$ and that this led to a significant 
difference in the number of pairs created by these two polarisations, with differences 
growing with external-field phase. To put
these results into context, since the photon polarisation is transverse to its
wavevector, and since the simulation was carried out well into the equilibrium region where 
initial seed acceleration was shown to make little difference, one might speculate that 
the photon wavevectors and hence polarisations are in general isotropic. However, when one 
takes into account the fact that those photon wavevectors in the negative $\hat{z}$ hemisphere
are more likely to be generated as Compton scattering in this direction is more 
likely, with radiation emitted in the $1/\gamma$ emission cone of a
relativistic fermion experiencing bremsstrahlung, and that these photons are
also more likely to lead to pair-creation, then, broadly speaking, 
the polarisations $\phi=0,\pi/2$ correspond to the $(0,1,0,0)$ and $(0,0,1,0)$ 
directions respectively. The simulations showed that the polarisation distribution tended 
to that of an average over single Compton-scattering events, with no smoothing from 
polarisation-dependent pair-creation. That the distribution did not itself evolve in time 
is a sign that the scattering rates of later generations of created pairs were not correlated 
with the polarisations of photons used to generate them. For this to be included, spin- and 
polarisation- dependent rates must at least be used, allowing correlations to be present over 
successive generations. To the current level of approximation, it was 
shown that polarisation can play an important role
if it can be modified by its environment between scattering and creation events.\\

\begin{figure}[!h]
\centering\noindent
 \includegraphics[draft=false, width=4cm]{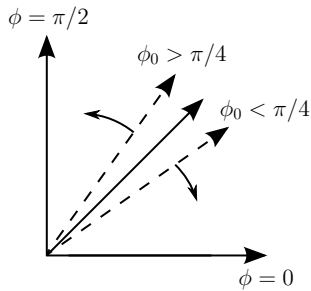}
\caption{For a constant phase shift in the $x$- and $y$- directions of a photon propagating in the $z$ direction, the diagram represents how the $\phi=\pi/4$ fixed-point is repulsive and the $\phi=0,\pi/2$ attractive (solid arrows) for initially random photon polarisations (dashed-line arrows). Analogous behaviour follows in the other quadrant $\pi/2<\phi<\pi$.}\label{fig:dichroism}
\end{figure}

%
It is a straightforward calculation to show that 
if a photon polarised in the $x$-$y$ plane with an initial angle $\phi_{0}$ to 
the $x$-axis gains a constant phase change $\delta\varphi_{x,y} \ll 1$ along
each of 
these axes, the subsequent rotation angle (dichroism) of the polarisation vector
becomes 
$\delta\phi = -[(\delta \varphi_{x}-\delta \varphi_{y})^{2}/8]\,\sin4\phi_{0}$, and the 
induced ellipticity is $\eps = [(\delta\varphi_{x}-\delta\varphi_{y})/2]\sin2\phi_{0}$. It follows 
that a randomly-aligned polarisation vector will eventually become either ordinary or extra-ordinary 
depending on $\phi_{0}$ and remain so (see \figref{fig:dichroism}). Likewise
the ellipticity will eventually become zero and 
the photon linearly-polarised. Although an idealistic model, the simulations presented in the 
current paper would imply a modification to pair-creation rates in such an
environment. \\

Phase changes can be induced by the polarised 
vacuum, where it has been shown in more complicated backgrounds such as e.g. focused lasers \cite{king10b}, 
that $\psi \gg \varepsilon$ is possible, but also in a plasma itself, 
in which it has been shown that vacuum polarisation effects can also be enhanced 
\cite{dipiazza07}.
Although likely irrelevant for laser-based experiments as the induced $\psi$ is too small to lead to a significant change in $N_{\gamma}$ \cite{king10b,dipiazza06,*heinzl06}, the polarisation-dependent results derived in 
the current paper could be of importance in astrophysical scenarios such as in the field around the
magnetospheres of pulsars where $\chi_{E}>0.1$ is possible (one example of this is in the 
soft gamma-ray repeater SGR 1900+14, \cite{kouveliotou93}). More thorough calculation and 
modelling of the strong magnetic field is required before the influence of these results in this area 
can be ascertained.

\section{Summary}
We have presented a derivation for the rate of nonlinear Compton scattering and
photon-seeded pair-creation for linearly-polarised photons in a constant crossed
field. Depending on the specific polarisation of the photon involved, the rate for Compton scattering (which also depends on photon frequency) is predicted to vary between $\pm 70\%$ and the rate for photon-seeded pair-creation $+35\%$, $-25\%$ that of the unpolarised rates for non-linear quantum parameters $\chi_{1} \in [0.1,10]$, $\chi_{k} \in [1,10]$ respectively. Moreover, 
those polarisations of photon that were more likely to be produced by nonlinear Compton scattering 
were less likely to be produced by pair-creation and vice versa. To study the combined effect of this 
disparity, the two-step electron-seeded pair-creation process $e^{\pm}\to e^{\pm} + \gamma$, $\gamma \to e^{+}e^{-}$ 
was approximated in a constant crossed field using tree-level rates integrated over lightfront momenta.
Analytical results show that when the photon has a \emph{fixed} polarisation, for incident electron chi-parameter
$1<\chi_{1}<100$ the rate for two-step electron-/positron- seeded pair creation can be around $15\%$ to $25\%$ lower 
than when the photon is considered unpolarised. However, the results also show that when 
the polarisation of the photon is averaged over, the difference from using unpolarised rates 
for each part of this two-step process, was only around $5\%$. 
To test whether photon polarisation plays a role in an ensemble and when other chains of
Compton-scattering and pair-creation events are involved, for example in the
creation of an electron-positron plasma, 
we used the numerical framework developed in \cite{ruhl10, elkina11}.
The results of simulations in a rotating electric field of frequency $1~\trm{eV}$ with
intensity-parameter $\xi=10^{4}$ for the two cases of having $10^{3}$ initial electrons with 
$\chi_{1} =0$ and $\chi_{1} = 5$ were shown to support these conclusions for electromagnetic
cascades of on average between two and three generations. On the one hand, the 
agreement to within $5\%$ in the number and spectrum of pairs created by using polarised and unpolarised photons 
was found, supporting this approximation when simulating electron-positron plasmas 
in intense lasers. On the other, simulations also agreed with another prediction from 
theory, that the difference between the most and least prevalent photon polarisations 
produced by one nonlinear Compton scattering event is more than $300\%$, with the difference 
in the number of pairs created from photons with these polarisations being more than
$40\%$. These results are particularly relevant when the photon vectors are anisotropic and when the 
photons' polarisation can be modified by its environment.

\section{Acknowledgements}
B. K. would like to acknowledge useful discussions with P. B\"ohl. This work was supported by the Grant No. DFG, FOR1048, RU633/1-1, by SFB TR18, projects B12 and B13 and by the Cluster-of-Excellence ‘‘Munich-Centre for Advanced Photonics’’ (MAP).

\bibliography{current}

\providecommand{\noopsort}[1]{}
\begin{thebibliography}{43}
\expandafter\ifx\csname natexlab\endcsname\relax\def\natexlab#1{#1}\fi
\expandafter\ifx\csname bibnamefont\endcsname\relax
  \def\bibnamefont#1{#1}\fi
\expandafter\ifx\csname bibfnamefont\endcsname\relax
  \def\bibfnamefont#1{#1}\fi
\expandafter\ifx\csname citenamefont\endcsname\relax
  \def\citenamefont#1{#1}\fi
\expandafter\ifx\csname url\endcsname\relax
  \def\url#1{\texttt{#1}}\fi
\expandafter\ifx\csname urlprefix\endcsname\relax\def\urlprefix{URL }\fi
\providecommand{\bibinfo}[2]{#2}
\providecommand{\eprint}[2][]{\url{#2}}

\bibitem[{\citenamefont{Bohr and Wheeler}(1939)}]{bohr39}
\bibinfo{author}{\bibfnamefont{N.}~\bibnamefont{Bohr}} \bibnamefont{and}
  \bibinfo{author}{\bibfnamefont{J.~A.} \bibnamefont{Wheeler}},
  \bibinfo{journal}{Phys. Rep.} \textbf{\bibinfo{volume}{56}},
  \bibinfo{pages}{426} (\bibinfo{year}{1939}).

\bibitem[{\citenamefont{Blackett and Occhialini}(1933)}]{blackett33}
\bibinfo{author}{\bibfnamefont{P.~M.~S.} \bibnamefont{Blackett}}
  \bibnamefont{and} \bibinfo{author}{\bibfnamefont{G.~P.~S.}
  \bibnamefont{Occhialini}}, \bibinfo{journal}{Proc. R. Soc. Lond. A}
  \textbf{\bibinfo{volume}{139}}, \bibinfo{pages}{699} (\bibinfo{year}{1933}).

\bibitem[{\citenamefont{Heitler}(1960)}]{heitler60}
\bibinfo{author}{\bibfnamefont{W.}~\bibnamefont{Heitler}},
  \emph{\bibinfo{title}{{The quantum theory of radiation (3rd edition)}}}
  (\bibinfo{publisher}{Oxford University Press}, \bibinfo{address}{Amen House,
  London E. C. 4}, \bibinfo{year}{1960}).

\bibitem[{\citenamefont{Bethe and Heitler}(1934)}]{bethe34}
\bibinfo{author}{\bibfnamefont{H.}~\bibnamefont{Bethe}} \bibnamefont{and}
  \bibinfo{author}{\bibfnamefont{W.}~\bibnamefont{Heitler}},
  \bibinfo{journal}{Proc. R. Soc. Lond. A} \textbf{\bibinfo{volume}{146}},
  \bibinfo{pages}{83} (\bibinfo{year}{1934}).

\bibitem[{\citenamefont{Nerush et~al.}(2011)}]{ruhl10}
\bibinfo{author}{\bibfnamefont{E.~N.} \bibnamefont{Nerush}}
  \bibnamefont{et~al.}, \bibinfo{journal}{Phys. Rev. Lett.}
  \textbf{\bibinfo{volume}{106}}, \bibinfo{pages}{035001}
  (\bibinfo{year}{2011}).

\bibitem[{\citenamefont{Elkina et~al.}(2011)}]{elkina11}
\bibinfo{author}{\bibfnamefont{N.~V.} \bibnamefont{Elkina}}
  \bibnamefont{et~al.}, \bibinfo{journal}{Phys. Rev. ST Accel. Beams}
  \textbf{\bibinfo{volume}{14}}, \bibinfo{pages}{054401}
  (\bibinfo{year}{2011}).

\bibitem[{\citenamefont{Ridgers et~al.}(2012)}]{ridgers12}
\bibinfo{author}{\bibfnamefont{C.~P.} \bibnamefont{Ridgers}}
  \bibnamefont{et~al.}, \bibinfo{journal}{Phys. Rev. Lett.}
  \textbf{\bibinfo{volume}{108}}, \bibinfo{pages}{165006}
  (\bibinfo{year}{2012}).

\bibitem[{\citenamefont{Burke et~al.}(1997)}]{burke97}
\bibinfo{author}{\bibfnamefont{D.~L.} \bibnamefont{Burke}}
  \bibnamefont{et~al.}, \bibinfo{journal}{Phys. Rev. Lett.}
  \textbf{\bibinfo{volume}{79}}, \bibinfo{pages}{1626} (\bibinfo{year}{1997}).

\bibitem[{\citenamefont{Chen et~al.}(2009)}]{chen09}
\bibinfo{author}{\bibfnamefont{H.}~\bibnamefont{Chen}} \bibnamefont{et~al.},
  \bibinfo{journal}{Phys. Rev. Lett.} \textbf{\bibinfo{volume}{102}},
  \bibinfo{pages}{105001} (\bibinfo{year}{2009}).

\bibitem[{\citenamefont{Hebenstreit et~al.}(2013)\citenamefont{Hebenstreit,
  Berges, and Gelfand}}]{hebenstreit13}
\bibinfo{author}{\bibfnamefont{F.}~\bibnamefont{Hebenstreit}},
  \bibinfo{author}{\bibfnamefont{J.}~\bibnamefont{Berges}}, \bibnamefont{and}
  \bibinfo{author}{\bibfnamefont{D.}~\bibnamefont{Gelfand}},
  \emph{\bibinfo{title}{Simulating fermion production in 1 + 1 dimensional
  qed}}, \bibinfo{howpublished}{\url{http://arxiv.org/abs/1302.5537}}
  (\bibinfo{year}{2013}).

\bibitem[{\citenamefont{Zhao et~al.}(2013)}]{ilderton13c}
\bibinfo{author}{\bibfnamefont{X.}~\bibnamefont{Zhao}} \bibnamefont{et~al.},
  \emph{\bibinfo{title}{Scattering in time-dependent basis light-front
  quantization}}, \bibinfo{howpublished}{\url{http://arxiv.org/abs/1303.3273}
  [nucl-th]} (\bibinfo{year}{2013}).

\bibitem[{\citenamefont{Brown and Kibble}(1964)}]{kibble64}
\bibinfo{author}{\bibfnamefont{L.~S.} \bibnamefont{Brown}} \bibnamefont{and}
  \bibinfo{author}{\bibfnamefont{T.~W.~B.} \bibnamefont{Kibble}},
  \bibinfo{journal}{Phys. Rep.} \textbf{\bibinfo{volume}{133}},
  \bibinfo{pages}{A705} (\bibinfo{year}{1964}).

\bibitem[{\citenamefont{Nikishov and Ritus}(1964)}]{nikishov64}
\bibinfo{author}{\bibfnamefont{A.~I.} \bibnamefont{Nikishov}} \bibnamefont{and}
  \bibinfo{author}{\bibfnamefont{V.~I.} \bibnamefont{Ritus}},
  \bibinfo{journal}{Sov. Phys. JETP} \textbf{\bibinfo{volume}{19}},
  \bibinfo{pages}{529} (\bibinfo{year}{1964}).

\bibitem[{\citenamefont{Krajewska and
  Kami{\'n}ski}(2012{\natexlab{a}})}]{krajewska12a}
\bibinfo{author}{\bibfnamefont{K.}~\bibnamefont{Krajewska}} \bibnamefont{and}
  \bibinfo{author}{\bibfnamefont{J.~Z.} \bibnamefont{Kami{\'n}ski}},
  \bibinfo{journal}{Phys. Rev. A} \textbf{\bibinfo{volume}{85}},
  \bibinfo{pages}{062102} (\bibinfo{year}{2012}{\natexlab{a}}).

\bibitem[{\citenamefont{Krajewska and
  Kami{\'n}ski}(2012{\natexlab{b}})}]{krajewska12b}
\bibinfo{author}{\bibfnamefont{K.}~\bibnamefont{Krajewska}} \bibnamefont{and}
  \bibinfo{author}{\bibfnamefont{J.~Z.} \bibnamefont{Kami{\'n}ski}},
  \bibinfo{journal}{Phys. Rev. A} \textbf{\bibinfo{volume}{86}},
  \bibinfo{pages}{052104} (\bibinfo{year}{2012}{\natexlab{b}}).

\bibitem[{\citenamefont{Ritus}(1985)}]{ritus85}
\bibinfo{author}{\bibfnamefont{V.~I.} \bibnamefont{Ritus}},
  \bibinfo{journal}{J. Russ. Laser Res.} \textbf{\bibinfo{volume}{6}},
  \bibinfo{pages}{497} (\bibinfo{year}{1985}).

\bibitem[{\citenamefont{Marklund and Shukla}(2006)}]{marklund_review06}
\bibinfo{author}{\bibfnamefont{M.}~\bibnamefont{Marklund}} \bibnamefont{and}
  \bibinfo{author}{\bibfnamefont{P.~K.} \bibnamefont{Shukla}},
  \bibinfo{journal}{Rev. Mod. Phys.} \textbf{\bibinfo{volume}{78}},
  \bibinfo{pages}{591} (\bibinfo{year}{2006}).

\bibitem[{\citenamefont{Ehlotzky et~al.}(2009)\citenamefont{Ehlotzky,
  Krajewska, and Kami{\'n}ski}}]{ehlotzky09}
\bibinfo{author}{\bibfnamefont{F.}~\bibnamefont{Ehlotzky}},
  \bibinfo{author}{\bibfnamefont{K.}~\bibnamefont{Krajewska}},
  \bibnamefont{and} \bibinfo{author}{\bibfnamefont{J.~Z.}
  \bibnamefont{Kami{\'n}ski}}, \bibinfo{journal}{Rep. Prog. Phys.}
  \textbf{\bibinfo{volume}{72}}, \bibinfo{pages}{046401}
  (\bibinfo{year}{2009}).

\bibitem[{\citenamefont{{Di Piazza} et~al.}(2012)}]{dipiazza12}
\bibinfo{author}{\bibfnamefont{A.}~\bibnamefont{{Di Piazza}}}
  \bibnamefont{et~al.}, \bibinfo{journal}{Rev. Mod. Phys.}
  \textbf{\bibinfo{volume}{84}}, \bibinfo{pages}{1177} (\bibinfo{year}{2012}).

\bibitem[{\citenamefont{Thompson and Duncan}(1995)}]{thompson95}
\bibinfo{author}{\bibfnamefont{C.}~\bibnamefont{Thompson}} \bibnamefont{and}
  \bibinfo{author}{\bibfnamefont{R.~C.} \bibnamefont{Duncan}},
  \bibinfo{journal}{Mon. Not. R. Astron. Soc.}
  \textbf{\bibinfo{volume}{{275}}}, \bibinfo{pages}{255}
  (\bibinfo{year}{1995}).

\bibitem[{\citenamefont{Harding and Lai}(2006)}]{harding06}
\bibinfo{author}{\bibfnamefont{A.~K.} \bibnamefont{Harding}} \bibnamefont{and}
  \bibinfo{author}{\bibfnamefont{D.}~\bibnamefont{Lai}}, \bibinfo{journal}{Rep.
  Prog. Phys.} \textbf{\bibinfo{volume}{69}}, \bibinfo{pages}{2631}
  (\bibinfo{year}{2006}).

\bibitem[{\citenamefont{Heinzl and Ilderton}(2009)}]{ilderton09}
\bibinfo{author}{\bibfnamefont{T.}~\bibnamefont{Heinzl}} \bibnamefont{and}
  \bibinfo{author}{\bibfnamefont{A.}~\bibnamefont{Ilderton}},
  \bibinfo{journal}{Opt. Commun.} \textbf{\bibinfo{volume}{282}},
  \bibinfo{pages}{1879} (\bibinfo{year}{2009}).

\bibitem[{\citenamefont{Dinu et~al.}(2012)\citenamefont{Dinu, Heinzl, and
  Ilderton}}]{dinu12}
\bibinfo{author}{\bibfnamefont{V.}~\bibnamefont{Dinu}},
  \bibinfo{author}{\bibfnamefont{T.}~\bibnamefont{Heinzl}}, \bibnamefont{and}
  \bibinfo{author}{\bibfnamefont{A.}~\bibnamefont{Ilderton}},
  \bibinfo{journal}{Phys. Rev. D} \textbf{\bibinfo{volume}{86}},
  \bibinfo{pages}{085037} (\bibinfo{year}{2012}).

\bibitem[{\citenamefont{Volkov}(1935)}]{volkov35}
\bibinfo{author}{\bibfnamefont{D.~M.} \bibnamefont{Volkov}},
  \bibinfo{journal}{Z. Phys.} \textbf{\bibinfo{volume}{94}},
  \bibinfo{pages}{250} (\bibinfo{year}{1935}).

\bibitem[{\citenamefont{Berestetskii et~al.}(1982)\citenamefont{Berestetskii,
  Lifshitz, and Pitaevskii}}]{landau4}
\bibinfo{author}{\bibfnamefont{V.~B.} \bibnamefont{Berestetskii}},
  \bibinfo{author}{\bibfnamefont{E.~M.} \bibnamefont{Lifshitz}},
  \bibnamefont{and} \bibinfo{author}{\bibfnamefont{L.~P.}
  \bibnamefont{Pitaevskii}}, \emph{\bibinfo{title}{Quantum Electrodynamics
  (second edition)}} (\bibinfo{publisher}{Butterworth-Heinemann},
  \bibinfo{address}{Oxford}, \bibinfo{year}{1982}).

\bibitem[{\citenamefont{Olver}(1997)}]{olver97}
\bibinfo{author}{\bibfnamefont{F.~W.~J.} \bibnamefont{Olver}},
  \emph{\bibinfo{title}{Asymptotics and Special Functions}}
  (\bibinfo{publisher}{AKP Classics}, \bibinfo{address}{A K Peters Ltd., 63
  South Avenue, Natick, MA 01760}, \bibinfo{year}{1997}).

\bibitem[{\citenamefont{Ba\u{\i}er et~al.}(1976)\citenamefont{Ba\u{\i}er,
  Mil'shte\u{\i}n, and Strakhovenko}}]{baier75a}
\bibinfo{author}{\bibfnamefont{V.~N.} \bibnamefont{Ba\u{\i}er}},
  \bibinfo{author}{\bibfnamefont{A.~I.} \bibnamefont{Mil'shte\u{\i}n}},
  \bibnamefont{and} \bibinfo{author}{\bibfnamefont{V.~M.}
  \bibnamefont{Strakhovenko}}, \bibinfo{journal}{Sov. Phys. JETP}
  \textbf{\bibinfo{volume}{42}}, \bibinfo{pages}{961} (\bibinfo{year}{1976}).

\bibitem[{\citenamefont{King and Ruhl}(2013)}]{king13b}
\bibinfo{author}{\bibfnamefont{B.}~\bibnamefont{King}} \bibnamefont{and}
  \bibinfo{author}{\bibfnamefont{H.}~\bibnamefont{Ruhl}},
  \emph{\bibinfo{title}{The trident process in a constant crossed field}},
  \bibinfo{howpublished}{\url{http://arxiv.org/abs/1303.1356} [hep-th]}
  (\bibinfo{year}{2013}).

\bibitem[{\citenamefont{Aspnes}(1966)}]{aspnes66}
\bibinfo{author}{\bibfnamefont{D.~E.} \bibnamefont{Aspnes}},
  \bibinfo{journal}{Phys. Rep.} \textbf{\bibinfo{volume}{147}},
  \bibinfo{pages}{554} (\bibinfo{year}{1966}).

\bibitem[{\citenamefont{Jackson}(1999)}]{jackson99}
\bibinfo{author}{\bibfnamefont{J.~D.} \bibnamefont{Jackson}},
  \emph{\bibinfo{title}{Classical Electrodynamics (3rd Edition)}}
  (\bibinfo{publisher}{John Wiley \& Sons, Inc.}, \bibinfo{address}{New York},
  \bibinfo{year}{1999}).

\bibitem[{\citenamefont{Harvey et~al.}(2009)\citenamefont{Harvey, Heinzl, and
  Ilderton}}]{harvey09}
\bibinfo{author}{\bibfnamefont{C.}~\bibnamefont{Harvey}},
  \bibinfo{author}{\bibfnamefont{T.}~\bibnamefont{Heinzl}}, \bibnamefont{and}
  \bibinfo{author}{\bibfnamefont{A.}~\bibnamefont{Ilderton}},
  \bibinfo{journal}{Phys. Rev. A} \textbf{\bibinfo{volume}{79}},
  \bibinfo{pages}{063407} (\bibinfo{year}{2009}).

\bibitem[{\citenamefont{Ilderton}(2011)}]{ilderton11}
\bibinfo{author}{\bibfnamefont{A.}~\bibnamefont{Ilderton}},
  \bibinfo{journal}{Phys. Rev. Lett.} \textbf{\bibinfo{volume}{106}},
  \bibinfo{pages}{020404} (\bibinfo{year}{2011}).

\bibitem[{\citenamefont{Hu et~al.}(2010)\citenamefont{Hu, M{\"u}ller, and
  Keitel}}]{hu10}
\bibinfo{author}{\bibfnamefont{H.}~\bibnamefont{Hu}},
  \bibinfo{author}{\bibfnamefont{C.}~\bibnamefont{M{\"u}ller}},
  \bibnamefont{and} \bibinfo{author}{\bibfnamefont{C.~H.}
  \bibnamefont{Keitel}}, \bibinfo{journal}{Phys. Rev. Lett.}
  \textbf{\bibinfo{volume}{105}}, \bibinfo{pages}{080401}
  (\bibinfo{year}{2010}).

\bibitem[{\citenamefont{Meuren and Di~Piazza}(2011)}]{meuren11}
\bibinfo{author}{\bibfnamefont{S.}~\bibnamefont{Meuren}} \bibnamefont{and}
  \bibinfo{author}{\bibfnamefont{A.}~\bibnamefont{Di~Piazza}},
  \bibinfo{journal}{Phys. Rev. Lett.} \textbf{\bibinfo{volume}{107}},
  \bibinfo{pages}{260401} (\bibinfo{year}{2011}).

\bibitem[{\citenamefont{Seipt and K\"ampfer}(2012)}]{seipt12}
\bibinfo{author}{\bibfnamefont{D.}~\bibnamefont{Seipt}} \bibnamefont{and}
  \bibinfo{author}{\bibfnamefont{B.}~\bibnamefont{K\"ampfer}},
  \bibinfo{journal}{Phys. Rev. D} \textbf{\bibinfo{volume}{85}},
  \bibinfo{pages}{101701} (\bibinfo{year}{2012}).

\bibitem[{\citenamefont{Mackenroth and Di~Piazza}(2013)}]{mackenroth13}
\bibinfo{author}{\bibfnamefont{F.}~\bibnamefont{Mackenroth}} \bibnamefont{and}
  \bibinfo{author}{\bibfnamefont{A.}~\bibnamefont{Di~Piazza}},
  \bibinfo{journal}{Phys. Rev. Lett.} \textbf{\bibinfo{volume}{110}},
  \bibinfo{pages}{070402} (\bibinfo{year}{2013}).

\bibitem[{\citenamefont{Fedotov et~al.}(2010)\citenamefont{Fedotov, Narozhny,
  Mourou, and Korn}}]{fedotov10}
\bibinfo{author}{\bibfnamefont{A.~M.} \bibnamefont{Fedotov}},
  \bibinfo{author}{\bibfnamefont{N.~B.} \bibnamefont{Narozhny}},
  \bibinfo{author}{\bibfnamefont{G.}~\bibnamefont{Mourou}}, \bibnamefont{and}
  \bibinfo{author}{\bibfnamefont{G.}~\bibnamefont{Korn}},
  \bibinfo{journal}{Phys. Rev. Lett.} \textbf{\bibinfo{volume}{105}},
  \bibinfo{pages}{080402} (\bibinfo{year}{2010}).

\bibitem[{\citenamefont{Neitz and Di~Piazza}(2013)}]{neitz13}
\bibinfo{author}{\bibfnamefont{N.}~\bibnamefont{Neitz}} \bibnamefont{and}
  \bibinfo{author}{\bibfnamefont{A.}~\bibnamefont{Di~Piazza}},
  \emph{\bibinfo{title}{Stochasticity effects in quantum radiation reaction}},
  \bibinfo{howpublished}{\url{http://arxiv.org/abs/1301.5524}}
  (\bibinfo{year}{2013}).

\bibitem[{\citenamefont{King et~al.}(2010)\citenamefont{King, {Di Piazza}, and
  Keitel}}]{king10b}
\bibinfo{author}{\bibfnamefont{B.}~\bibnamefont{King}},
  \bibinfo{author}{\bibfnamefont{A.}~\bibnamefont{{Di Piazza}}},
  \bibnamefont{and} \bibinfo{author}{\bibfnamefont{C.~H.}
  \bibnamefont{Keitel}}, \bibinfo{journal}{Phys. Rev. A}
  \textbf{\bibinfo{volume}{82}}, \bibinfo{pages}{032114}
  (\bibinfo{year}{2010}).

\bibitem[{\citenamefont{{Di Piazza} et~al.}(2007)\citenamefont{{Di Piazza},
  Hatsagortsyan, and Keitel}}]{dipiazza07}
\bibinfo{author}{\bibfnamefont{A.}~\bibnamefont{{Di Piazza}}},
  \bibinfo{author}{\bibfnamefont{K.~Z.} \bibnamefont{Hatsagortsyan}},
  \bibnamefont{and} \bibinfo{author}{\bibfnamefont{C.~H.}
  \bibnamefont{Keitel}}, \bibinfo{journal}{Phys. Plasmas}
  \textbf{\bibinfo{volume}{14}}, \bibinfo{pages}{032102}
  (\bibinfo{year}{2007}).

\bibitem[{\citenamefont{{Di Piazza} et~al.}(2006)\citenamefont{{Di Piazza},
  Hatsagortsyan, and Keitel}}]{dipiazza06}
\bibinfo{author}{\bibfnamefont{A.}~\bibnamefont{{Di Piazza}}},
  \bibinfo{author}{\bibfnamefont{K.~Z.} \bibnamefont{Hatsagortsyan}},
  \bibnamefont{and} \bibinfo{author}{\bibfnamefont{C.~H.}
  \bibnamefont{Keitel}}, \bibinfo{journal}{Phys. Rev. Lett.}
  \textbf{\bibinfo{volume}{97}}, \bibinfo{pages}{083603}
  (\bibinfo{year}{2006}).

\bibitem[{\citenamefont{Heinzl et~al.}(2006)}]{heinzl06}
\bibinfo{author}{\bibfnamefont{T.}~\bibnamefont{Heinzl}} \bibnamefont{et~al.},
  \bibinfo{journal}{Opt. Commun.} \textbf{\bibinfo{volume}{267}},
  \bibinfo{pages}{318} (\bibinfo{year}{2006}).

\bibitem[{\citenamefont{Kouveliotou et~al.}(1993)}]{kouveliotou93}
\bibinfo{author}{\bibfnamefont{C.}~\bibnamefont{Kouveliotou}}
  \bibnamefont{et~al.}, \bibinfo{journal}{Nature}
  \textbf{\bibinfo{volume}{362}}, \bibinfo{pages}{728} (\bibinfo{year}{1993}).

\end{thebibliography}

\end{document}